\documentclass[11pt]{article}
\usepackage{amsmath,amsfonts,amssymb}
\usepackage{graphicx,color}
\usepackage{geometry}
\usepackage{hyperref}
\usepackage{dutchcal}
\usepackage{authblk}

\newcommand{\dif}{{\mathrm{d}}}

\newcommand{\Exp}{\mathbb{E}}

\newcommand{\bcdot}{\boldsymbol{\cdot}}

\title{Stochastic Transport and Wave Interactions for Multiscale Surface Gravity Waves: Part II: Kinetic Theory and Ocean-Wave Applications}

\date{}
\author[1]{E. M\'emin}
\author[2,5]{B. Chapron}
\author[3,4]{A. Debussche}
\author[5]{L. Mari\'e}

\affil[1]{Univ Rennes, Inria, Odyssey, IRMAR-UMR 6625, Centre Inria Rennes 35042 Rennes Cedex, France}
\affil[2]{Ifremer, Inria, Odyssey, F-29280 Plouzan\'e, France}
\affil[3]{Univ Rennes, CNRS, IRMAR-UMR 6625, F-35000 Rennes, France}
\affil[4]{Institut universitaire de France (IUF)}
\affil[5]{Univ Brest, Ifremer, CNRS, IRD, Laboratoire d'Oc\'eanographie Physique et Spatiale (LOPS), F-29280 Plouzan\'e, France}

\begin{document}
\maketitle

\begin{abstract}
Building on the stochastic variational framework established in the companion paper,
we investigate here the linearized stochastic water-wave system, consisting of a
large-scale stochastic wave dynamics coupled to transport dynamics for the
small-scale correlation modes.
Within this framework, we develop, in the deep-water regime,  a kinetic theory for surface gravity waves interacting
with unresolved stochastic velocity fields.

An energy analysis yields a wave-action kinetic equation exhibiting two distinct
regimes: a diffusive scattering regime and a quartic interaction regime with structural
similarities to Hasselmann--Zakharov theory.
In the present framework, these effective quartic interactions arise through stochastic
transport of unresolved fluctuations by the large-scale flow rather than through
classical intrinsic resonant nonlinearity.
Scaling laws are derived for the diffusion tensor and the effective growth rate,
revealing a Miles-type production--dissipation mechanism.

Using JONSWAP spectra, we then compare the strength of stochastic transport and
classical Hasselmann interactions.
For realistic oceanic values of unresolved velocity variance
($\sigma_u \approx 0.1\,\mathrm{m\,s^{-1}}$)
and decorrelation time
($\tau_c \approx 10\,\mathrm{s}$),
stochastic transport is found to compete with, and often exceed,
classical four-wave interaction rates over broad spectral ranges.
The transport intensity $S=\sigma_u^2\tau_c$
emerges as a key parameter controlling the transition between interaction regimes.

These results suggest that unresolved stochastic transport may play a substantially
larger role in spectral evolution than is commonly represented in operational wave models,
and motivate the inclusion of transport-induced source terms alongside standard
resonant interaction closures.
\end{abstract}

\section{Introduction}

The companion paper~\cite{DebusscheMemin2026a} developed a stochastic variational framework for surface gravity waves in which unresolved scales enter through a regularized stochastic transport acting on the velocity potential. Starting from Luke's variational principle and a decomposition into resolved and unresolved components, that work led to a coupled multiscale system governed by two intertwined principles: a pathwise variational principle for the large-scale wave dynamics and a variational principle in expectation governing the evolution of the noise modal  functions. The resulting formulation preserves the Hamiltonian Zakharov--Craig--Sulem structure while providing a systematic framework for stochastic reduced models of ocean waves.

In this second paper, we investigate, for the deep-water regime, the physical consequences of that framework from a kinetic and modeling perspective. In particular, we examine how stochastic transport induced by unresolved motions modifies spectral wave evolution, and how this mechanism compares with the classical weak-turbulence picture embodied in the Hasselmann--Zakharov equation \cite{hasselmann1962nonlinear,hasselmann1963nonlinear,zakharov1966turbulence,zakharov1967weak,zakharov1968stability,zakharov1992kolmogorov}.

This question is motivated by a long-standing gap in wave modeling. Resonant four-wave interactions form a cornerstone of spectral wave theory, yet transport by unresolved currents, turbulence, internal waves, and submesoscale motions is generally represented only indirectly or neglected, despite classical physical arguments---notably the Kraichnan--Tennekes sweeping picture---suggesting that random advection may strongly influence decorrelation and spectral transfer. The stochastic variational framework of the companion paper provides a natural setting in which this mechanism can be studied systematically.

The present work has two main objectives:
\begin{enumerate}
\item \textbf{Develop a kinetic theory for stochastic transport.}
Starting from the linearized stochastic water-waves equations, we derive a wave-action kinetic equation exhibiting two complementary regimes: a diffusive scattering regime and a quartic interaction regime structurally reminiscent of the Hasselmann--Zakharov equation. In contrast to classical weak-turbulence theory, these effective quartic interactions arise here from stochastic transport by unresolved motions rather than intrinsic resonant nonlinearity.

\item \textbf{Quantify the relative importance of stochastic transport and Hasselmann--Zakharov four-waves resonant interactions.}
To compare these mechanisms, we introduce the dimensionless transport-to-resonance interaction ratio
\begin{equation}
\Gamma
=
\frac{k^2 \sigma_u^2 \tau_c}{\epsilon_\mu^4 \omega_k},
\end{equation}
where $\sigma_u^2$ denotes unresolved velocity variance, $\tau_c$ its decorrelation time, $\epsilon_\mu$ the wave steepness\footnote{The subscript $\mu$, the squared aspect ratio, indicates the relation between the nonlinearity parameter, $\epsilon$, used in the first paper and wave steepness}, and $\omega_k$ the wave frequency. Values $\Gamma>1$ correspond to transport-dominated evolution.
\end{enumerate}

Using JONSWAP spectral parameterizations, we evaluate this ratio for different sea-states and over realistic oceanic ranges of unresolved transport (eddy viscosity) intensity 
\[
S=\sigma_u^2\tau_c.
\]
Our analysis suggests that, for physically realistic values of $S$, stochastic transport can compete with, and often significantly exceed, classical four-wave interaction rates over broad spectral ranges. The results indicate that unresolved transport may play a substantially larger role in spectral evolution than is commonly represented in operational wave models.

These findings motivate a re-examination of spectral wave closures in which unresolved transport acts as an explicit source of spectral redistribution, alongside resonant nonlinear interactions. Beyond the kinetic theory itself, the analysis developed here yields implications for peak dynamics, swell decay, and wave forecasting, while also suggesting new parameterizations for stochastic source terms in operational spectral models.

The remainder of the paper is organized as follows. Section~2 recalls the stochastic water-wave formulation developed in the companion paper. Section~3 derives wave-action  kinetic equation, for the linear stochastic water-wave equation with prescribed (stationnary) noise. Its diffusive continuous limit is inferred together with the effects of forcing and dissipation. Section~4 develops the quartic kinetic interaction term associated to the complete linear stochastic water-wave system. Section~5 presents the comparison with JONSWAP spectra and the resulting regime analysis. Section~6 discusses implications for wave modeling and concludes.

\section{Stochastic water-wave formulation}

In this section we briefly recall the stochastic water-wave model derived in the companion paper \cite{DebusscheMemin2026a}; see also \cite{Debussche-Memin25}. The model is based on two coupled variational principles. The first is a pathwise variational principle governing the evolution of the large-scale velocity potential $\overline{\phi}$, while the second is a variational principle in expectation describing the dynamics of the unresolved small-scale component $\phi^\tau$.
The unresolved component is represented as a time-correlated random field through the modal decomposition
\begin{equation}
\phi^\tau(x,t)
=
\int_{t-\tau}^{t+\tau}
h_\tau(t-s)\,
\phi_i(x)\,
\mathrm d\beta_s^i,
\end{equation}
where $h_\tau$ is a family of smooth regularization kernels with compact support and unit integral, and the functions $\phi_i$ denote velocity-potential modes with bounded variance.

The full velocity potential is decomposed as
\begin{equation}
\label{W-v}
\phi(x,t)
=
\overline{\phi}(x,t)
+
\phi^\tau(x,t).
\end{equation}
Because the temporal regularization is symmetric in past and future times, the decorrelation limit $\tau\to0$ formally yields a Stratonovich stochastic transport through a Wong--Zakai type limit.
The regularized noise process
\begin{equation}
W_t^\tau
=
\int_{t-\tau}^{t+\tau}
h_\tau(t-s)\,
\mathrm d\beta_s
\end{equation}
has, under the kernel normalization considered here, variance $\mathbb E[(W_t^\tau)^2] =\tau^{-1}$,
so that $W_t^\tau=\mathcal O(\tau^{-1/2})$.
This regularization therefore represents a large but finite separation of temporal scales, rather than idealized white-in-time forcing.

The free-surface elevation satisfies the stochastic kinematic boundary condition expressing transport of particles along the moving interface. Throughout, we assume the free surface remains sufficiently smooth and exclude wave breaking.
Consistent with the decorrelation limit, the surface elevation is modeled as a semimartingale,
\begin{equation}
\mathrm d\eta
=
\widetilde{\eta}\,\mathrm dt
+
\mathrm d\eta^\tau.
\end{equation}
Although $\eta$ is stochastic, it remains significantly more regular than the velocity field in the decorrelation limit. In contrast to the velocity, which involves, at the limit, a rapidly fluctuating white-in-time component, the surface elevation retains enough regularity for evaluations at the free surface  to remain well-defined.

\subsection{Stochastic water-wave system}
The water waves dynamics involves a classical kinematic condition for the free surface $\eta$:
\begin{equation}
\partial_t \eta  = G[\eta] (\overline \varphi + \varphi^{\tau}),
\label{Dyn-Eta}
\end{equation}
where we introduced the Dirichlet-to-Neumann operator \cite{Craig-Sulem93,zakharov1968stability}:
\begin{equation}
G[\eta] \varphi = (\partial_z \phi - \nabla_{\!x} \phi\bcdot\nabla\eta)_{|_{z=\eta}}.
\label{def-DtoN}
\end{equation}
This operator maps a Dirichlet boundary value $\varphi$ of the velocity potential  to its normal derivative  at the free surface, thereby justifying its name. The properties of this operator has been extensively studied in the literature (see in particular \cite{Lannes-2013} and references therein). 

A pathwise variational principle provides the evolution of the trace at the surface of the large-scale velocity potential:
\begin{multline}
\label{Dyn-LS-Varphi}
\partial_t \overline{\varphi} = -\biggl( g\eta
+ \frac{1}{2} |\nabla \overline{\varphi}|^2
- \frac{1}{2}
\frac{\bigl(G[\eta]\overline{\varphi} + \nabla \eta \cdot \nabla \overline{\varphi}\bigr)^2}
{1+|\nabla\eta|^2}\\
\quad\qquad\qquad
+ \nabla \overline{\varphi} \cdot \nabla \varphi^\tau
- \frac{
\bigl(G[\eta]\overline{\varphi} + \nabla\eta \cdot \nabla \overline{\varphi}\bigr)
\bigl(G[\eta]\varphi^\tau + \nabla\eta \cdot \nabla \varphi^\tau\bigr)
}{1+|\nabla\eta|^2}\biggr),
\end{multline}
These evolution equations can be shown to conserve a Hamiltonian structure 
derived from a stochastic extension of the Craig--Sulem--Zakharov formulation, 
so that the resolved dynamics remains consistent with the underlying variational structure.

As for the unresolved component, its dynamics follows from the coupled variational principle in expectation associated with the Euler equations. The evolution of the correlation functions of the trace of the velocity potential at the free surface reads
\begin{equation}
\partial_t \varphi_i 
+ \nabla\overline{\varphi}\bcdot \nabla\varphi_i
-  \frac{(\nabla\varphi_i\bcdot\nabla\eta)
\bigl(G[\eta]\overline \varphi 
+ \nabla \overline{\varphi}\bcdot \nabla\eta\bigr)}
{1+ |\nabla\eta|^2}
-  \frac{
\bigl(G[\eta]\varphi_i + \nabla \varphi_i \bcdot \nabla\eta \bigr)
G[\eta] \varphi^\tau}
{1+ |\nabla\eta|^2}
=0.
\label{varphi-i-sto}
\end{equation}

Equations above together with the free-surface transport, define a coupled stochastic water-wave system for the resolved free surface dynamics  and the unresolved correlation modes.

In the present work we focus on a simplified regime 
obtained by linearizing the large-scale dynamics. In the following section we first examine a frozen-correlation approximation for the noise modes, in which the functions $\varphi_i$ are treated as prescribed and stationnary. In this first regime, unresolved modes act as a stochastic transport forcing on the resolved waves, rather than as an additional weakly nonlinear wave field; this is the central distinction explored in this paper.

  \section{Linearization of the large-scale water-wave system under fixed noise}
  In the following we provide in Fourier form the solution of the water-wave system linearized around flat surface at rest. 
  \begin{align}
    &\partial_t\overline{\varphi}+\eta+\nabla\varphi^\tau\bcdot\nabla\overline{\varphi}=0,\label{Linear-Large-scale}\\
    &\partial_t \eta =  \partial_z \phi_{|_{z=0}} +  \; \partial_z \phi^\tau_{|_{z=0}}.
    \label{eq-linear-S-Saint-Venant}
\end{align}
In nondimensional form, the advection of the large-scale velocity potential by the noise in  \eqref{Linear-Large-scale} -- referred to as transport-noise -- scales in $\epsilon_\sigma {\mathcal m}/\nu$, where the noise nonlinear parameter $\epsilon_\sigma = A_\sigma/H$ denotes the ratio between the characteristic noise wave height and finite water depth, while $\nu=\tanh(2\pi\sqrt\mu)/2\pi\sqrt{\mu}$ interpolates between Shallow-water and deep-water regimes, with $\mu=H^2/L^2$ \cite{DebusscheMemin2026a}. In deep-water, (for finite depth) $\nu=(2\pi\sqrt \mu)^{-1}$, which corresponds to the regime explored in this study, the transport-noise term $\nabla\varphi^\tau\bcdot\nabla\overline{\varphi}$ scales equivalently in terms of the slope $\epsilon^\sigma_\mu=A_\sigma/L$ as $\epsilon_\mu^\sigma \mathcal{m}$.     

 In this first case, the noise is assumed to be given and stationary. We will also assume lateral periodic conditions. As such, the noise can be specified in terms of Fourier basis functions as
 \begin{equation} 
  \varphi^\tau = \sum_j c_{k_j} e^{ik_j\bcdot x}\,  W_j^{\tau}(t) \text{ with } W_j^\tau(t) = \int^{t+\tau}_{t-\tau}\!\!\!h_\tau(t-s) \dif \beta^j_s.
  \end{equation}
  The wavenumbers \(k_j\) label the unresolved spatial components,
whereas \(k\) denotes a resolved-wave wavenumber. The coefficients
\(c_{k_j}\) are the Fourier coefficients of the prescribed
velocity-potential correlation modes and are time independent in the
present frozen-correlation approximation. They do not depend neither on the
regularization scale \(\tau\). Since
\(W_j^\tau=\mathcal O(\tau^{-1/2})\), dimensional considerations, with the noise velocity potential $\varphi^\tau \sim L^2/T$,  give
\(c_{k_j}\sim |k_j|^{-2}\sqrt{\omega_j}\), where
\(\omega_j=\omega(k_j)\) is used only to prescribe the wavenumber
scaling of the spatial covariance. No intrinsic temporal Airy carrier
is assigned to the noise modes in this first regime. For simplicity, the scaling of the transport-noise is assumed to be absorbed in the noise amplitude coefficient $c_{k_j}$.
  
  The transport-noise term in the linearized large-scale expression \eqref{Linear-Large-scale} reads:
  \begin{equation}
  \nabla \overline\varphi \bcdot \nabla \varphi^\tau = \sum_{k,j} -c_{k_j}(k-k_j)\bcdot k_j  \overline\varphi_{k-k_j} W^\tau_j(t) e^{i k\bcdot x}, 
  \end{equation}
  For a fixed horizontal wavenumber ${k}$,  the linearized water system around the flat surface $z=0$ is expressed in vectorial form in terms of $U_k=(\varphi_k,\eta_k)^{tr}$  as:
\begin{equation}
\partial_t U_k
=
A_{{k}} U_k
+ \sum_j W_j^\tau(t) B_{k,k_j} U_{k-k_j} +  F_{k}(t), 
\end{equation}
where 
\begin{equation}
A_k=
\left(\begin{matrix}
0 &-g \\
G_k[0] &0
\end{matrix}\right), 
B_{k,k_j}=\begin{pmatrix}
c_{k_j}\,k_j\bcdot(k-k_j) & 0\\
0& 0
\end{pmatrix},
F_{k}= 
\begin{pmatrix}
0 \\
\sum_j W_j^\tau(t) G_k[0] c_{k_j} \delta_{k,k_j}
\end{pmatrix},
\end{equation}
denotes respectively, the Airy matrix, a convolution multiplicative operator and an additive stochastic forcing applying only at noise modes as seen from the Kronecker $\delta$ function. 

Denoting the semi-group $S_k(t)=e^{tA_k}$ the exact mild (Duhamel) solution reads 
\begin{equation}
U_k(t) = S_k(t)U_k(0) + \sum_j \int_0^t S_k(t-s) B_{k,k_j} U_{k-k_j}(s) W_j^{\tau} (s) \dif s + \int_0^t S_k(t-s) F_k (s) \dif s.
\end{equation}
The Dirichlet-to-Neumann at the surface at rest, is given through Fourier multipliers by
\[
G_{{k}}[0] = |{k}| \tanh\!\left(|{k}| H\right).
\]
 The eigenvalues of matrix $A$ are $\lambda_\pm= \pm i \omega_k$, with eigenvectors $\boldsymbol{v}^{\pm}
=(\dfrac{\pm i\omega}{G_k[0]}, 1)^{tr}$ and the  dispersion relation
 \begin{equation}
 \omega_{\mathcal k}^2 =gG_k[0] = g|k|\tanh (|k|H).
 \end{equation}
 Noting that even power of the Airy matrix are $A^{2n} = (-\omega^2)^n\mathbb{I}_2 $, we obtain integral explicit expressions of the solution with: 
 \begin{equation}
 S_k(t) = \begin{pmatrix} \cos(\omega_kt ) &-\frac{g}{\omega_k}\sin(\omega_k t) \\  \frac{G_k[0]}{\omega_k}\sin(\omega_k t) &\cos(\omega_k t)\end{pmatrix}.
 \end{equation}
 Through normal mode expression of the deterministic solution, we obtain $\eta_k = a_k +a^*_{-k}$, $\overline\varphi_k=-\frac{i\omega_k}{G_k[0]}(a_k-a_{-k}^*)$, and the Airy variable
 \begin{equation}
 a_k = \frac{1}{2}(\eta_k + \frac{iG_k[0]}{\omega_k}\overline\varphi_k) = \frac{1}{2}(\eta_k + \frac{i\omega_k}{g}\overline\varphi_k).
 \end{equation}
 Note that usually a rescaling  $\widetilde{a}_k=\frac{\sqrt g}{\sqrt \omega}a_k$ is performed so that $\omega |\widetilde a_k|^2$ scales as an energy per horizontal area with dimension $L^3/T^2$. The modulus $|\widetilde a_k|^2$, denoted by $N_k$, is related to kinetic energy through $N_k=E_k/\omega$ and is referred to as wave action density. 
 
 Taking the time derivative and substituting the expressions of  $\varphi$ and $\eta$, we obtain
 \begin{equation}
 \partial_t a_k = -i\omega a_k +\frac{1}{2} \sum_j{\cal C}_{k,k_j} (a_{k-k_j} - a^*_{-(k-k_j)})W^\tau_j(t) +f_k,
 \end{equation}
 with 
 \begin{equation}
 {\cal C}_{k,k_j}  = c_{k_j} k_j \bcdot (k-k_j)\frac{G_k[0]}{\omega_k} \frac{\omega_{k-k_j}}{G_{k-k_j}[0]} \text{ and } f_k = \sum_j G_k[0] c_{k_j} W^\tau_j(t) \delta_{k,k_j}.
 \end{equation}
 Introducing the interaction representation $b_k(t)=a_k(t)e^{i\omega_k t}$, the term involving $a^*_{-(k-k_j)}$ oscillates at the sum frequency $\omega_k+\omega_{k-k_j}$ and is non-resonant.
Under the rotating-wave (or resonant) approximation its contribution averages out on slow kinetic timescales and is neglected.
We obtain:
 \begin{equation}
 \partial_t a_k = -i\omega_k a_k +\frac{1}{2} \sum_j{\cal C}_{k,k_j} a_{k-k_j}W^\tau_j(t) +f_k.
 \end{equation}
 A Duhamel solution reads then 
 \begin{equation}
 a_k = e^{-i\omega_k t } a_k (0) + \sum_j \int_0^t e^{-i\omega_k (t-s) }[\frac{1}{2} {\cal C}_{k,k_j} \,a_{k-k_j}\,W^\tau_j(s) +f_k(s)]\dif s.
 \label{D-sol-a}
 \end{equation}
 Equation \eqref{D-sol-a} has the structure of a random linear scattering equation,
with multiplicative transport-noise generating mode coupling
and additive forcing injecting energy at unresolved modes.

We obtain for the surface elevation 
\begin{equation}
\eta(x,t) = -\sum_k e^{ik\bcdot x -i\omega_k t } a_k (t) + e^{ik\bcdot x +i\omega_k t } a^*_{-k} (t)
\end{equation}
 and for the velocity potential 
 \begin{equation}
\overline\varphi(x,t) = \sum_k \frac{i\omega_k}{G_k[0]}[e^{ik\bcdot x -i\omega_k t } a_k (t) + e^{ik\bcdot x +i\omega_k t } a^*_{-k} (t)].
\end{equation}
This representation describes linear Airy waves coupled through a stochastic mode-scattering operator induced by transport-noise, together with an additive forcing acting at the unresolved noise wavenumbers. It provides the random scattering dynamics underlying the kinetic description developed in the next section.
\subsection{Kinetic equation for the linear stochastic water-wave dynamics with fixed noise}

In the following, we separate the transport-noise term from the additive
forcing. In the scattering regime established below, the former redistributes
wave action among the resolved modes, while the latter corresponds to wave-action
injection at the noise wavenumbers. We therefore first drop the additive-noise
contribution and retain both terms of the multiplicative coupling:
\begin{equation}
\partial_t a_k
=
-i\omega_k a_k
+\frac{1}{2}\sum_j {\cal C}_{k,k_j}
\left(a_{k-k_j}-a^*_{-(k-k_j)}\right)W_j^\tau(t).
\label{amplitude-fixed-noise}
\end{equation}
Here and below, we understand the conventional action normalization
$a_k^{\rm A}=\sqrt{g/\omega_k}\,a_k$ to have been absorbed into the
definition of the amplitude. Correspondingly, the factor
$\sqrt{\omega_{k-k_j}/\omega_k}$ is absorbed into
${\cal C}_{k,k_j}$. We retain the same notation to avoid overloading the
formulas.

In this analysis, we replace the smoothed noise $W_j^\tau$ by independent
Ornstein--Uhlenbeck (OU) processes $Z_j$. Since the closure derived below
depends only on the two-time covariance of the noise, this amounts to
approximating the covariance generated by the smoothing kernel $h^\tau$ by an
exponential covariance. For generality, we allow the correlation time to
depend on the mode:
\begin{equation}
\label{OU}
\dif Z_j=-\gamma_j Z_j\,\dif t+\sigma_j\,\dif B_j(t),
\end{equation}
where the $B_j$ are independent standard Brownian motions. We assume that the
OU processes are initialized in their stationary distributions, so that
\begin{equation}
\Exp[Z_j(t)Z_{j'}(s)]
=
\delta_{jj'}\frac{\sigma_j^2}{2\gamma_j}
e^{-\gamma_j|t-s|}.
\label{OU-covariance}
\end{equation}
Since $W_j^\tau$, and hence $Z_j$, scales as $T^{-1/2}$, both $\gamma_j$
and $\sigma_j$ have dimensions of frequency. Matching a variance of order
$\tau^{-1}$ and a correlation time of order $\tau$ gives
$\gamma_j=\mathcal O(\tau^{-1})$ and
$\sigma_j=\mathcal O(\tau^{-1})$.

We now remove the fast oscillations by defining
\begin{equation}
b_k(t)=a_k(t)e^{i\omega_k t}.
\end{equation}
With this change of variables, the evolution equation becomes
\begin{equation}
\partial_t b_k
=
\frac{1}{2}\sum_j {\cal C}_{k,k_j}b_{k-k_j}Z_j(t)
e^{i\Omega_{k,-k_j}t}
-\frac{1}{2}\sum_j {\cal C}_{k,k_j}b^*_{-(k-k_j)}Z_j(t)
e^{i\Omega_{k,k_j}t},
\label{b-k-evol}
\end{equation}
where
\begin{equation}
\Omega_{k,-k_j}=\omega_k-\omega_{k-k_j},
\qquad
\Omega_{k,k_j}=\omega_k+\omega_{k-k_j}.
\end{equation}
With the action normalization specified above, this transformation preserves
the modal wave action
\begin{equation}
N_k=\Exp\left(b_kb_k^*\right).
\end{equation}
Its evolution is given by
\begin{align}
\partial_tN_k
&=
\Exp\left((\partial_tb_k)b_k^*\right)
+\Exp\left(b_k\partial_tb_k^*\right),
\nonumber\\
&=
\frac{1}{2}\sum_j
\Exp\left(
{\cal C}_{k,k_j}Z_jb_{k-k_j}b_k^*
e^{i\Omega_{k,-k_j}t}
\right)
-\frac{1}{2}\sum_j
\Exp\left(
{\cal C}_{k,k_j}Z_jb^*_{-(k-k_j)}b_k^*
e^{i\Omega_{k,k_j}t}
\right)
+\mathrm{c.c.}
\label{E-evol}
\end{align}
The right-hand side does not vanish because the wave amplitudes $b$ become
correlated with the processes $Z_j$.

We assume (i) weak coupling, so that ${\cal C}_{k,k_j}$ is small, and
(ii) a separation of time scales in which $b$ evolves slowly compared with
the OU correlation time. Under these assumptions, the leading contribution
to $\partial_tN_k$ arises at second order in the coupling. The Duhamel
representation of \eqref{b-k-evol} is
\begin{multline}
b_k(t)
=
b_k(0)
+\frac{1}{2}\sum_j {\cal C}_{k,k_j}
\int_0^t Z_j(s)b_{k-k_j}(s)e^{i\Omega_{k,-k_j}s}\,\dif s
\\
-\frac{1}{2}\sum_j {\cal C}_{k,k_j}
\int_0^t Z_j(s)b^*_{-(k-k_j)}(s)e^{i\Omega_{k,k_j}s}\,\dif s,
\label{Duhamel-b}
\end{multline}
and similarly
\begin{multline}
b_k^*(t)
=
b_k^*(0)
+\frac{1}{2}\sum_j {\cal C}_{k,k_j}^*
\int_0^t Z_j(s)b^*_{k-k_j}(s)e^{-i\Omega_{k,-k_j}s}\,\dif s
\\
-\frac{1}{2}\sum_j {\cal C}_{k,k_j}^*
\int_0^t Z_j(s)b_{-(k-k_j)}(s)e^{-i\Omega_{k,k_j}s}\,\dif s.
\label{DSb*}
\end{multline}

We first consider the contribution involving the frequency difference
$\Omega_{k,-k_j}$. At leading order in the weak-coupling expansion, the
initial wave amplitudes and the noise are independent, and the amplitudes vary
slowly over the OU correlation time. Inserting \eqref{DSb*} into the first
expectation of \eqref{E-evol} therefore gives
\begin{equation}
\Exp\left[Z_j(t)b_{k-k_j}(t)b_k^*(t)\right]
\approx
\frac{1}{2}{\cal C}_{k,k_j}^*N_{k-k_j}(t)
\times
\int_0^t
\Exp\left[Z_j(t)Z_j(s)\right]
e^{-i\Omega_{k,-k_j}s}\,\dif s.
\label{normal-correlation}
\end{equation}
After multiplication by the external factors appearing in \eqref{E-evol}, we obtain
\begin{multline}
\frac{1}{2}
\Exp\left(
{\cal C}_{k,k_j}Z_j(t)b_{k-k_j}(t)b_k^*(t)
e^{i\Omega_{k,-k_j}t}
\right)
\approx
\frac{1}{4}|{\cal C}_{k,k_j}|^2N_{k-k_j}(t)
\\
\times
\int_0^t
\Exp\left[Z_j(t)Z_j(s)\right]
e^{i\Omega_{k,-k_j}(t-s)}\,\dif s.
\label{term1}
\end{multline}
Setting $u=t-s$ and using stationarity yields a memory integral depending
only on the time lag. In the long-time regime $t\gg\gamma_j^{-1}$,
\begin{equation}
{\cal R}_e\int_0^\infty R_j(u)e^{i\Omega_{k,-k_j}u}\,\dif u
=
\frac{\sigma_j^2}{2}
\frac{1}{\gamma_j^2+\Omega_{k,-k_j}^2}
=:\frac{1}{2}\widehat{R}_j(\Omega_{k,-k_j}),
\label{Rj-coef}
\end{equation}
where
\begin{equation}
R_j(u)=\frac{\sigma_j^2}{2\gamma_j}e^{-\gamma_j u},
\qquad u\geq0.
\end{equation}
Including the complex-conjugate contribution, we define the incoming
scattering rate
\begin{equation}
{\mathcal G}_{k,k_j}
=
\frac{1}{4}|{\cal C}_{k,k_j}|^2\widehat{R}_j(\Omega_{k,-k_j}).
\label{scattering-kernel}
\end{equation}
The coefficient ${\mathcal G}_{k,k_j}$ is non-negative and has dimension
$T^{-1}$.

We now return to the second expectation in \eqref{E-evol}. It involves the
anomalous correlation
\begin{equation}
\Exp\left[Z_j(t)b^*_{-(k-k_j)}(t)b_k^*(t)\right]
\end{equation}
and is weighted by the sum frequency
$\Omega_{k,k_j}=\omega_k+\omega_{k-k_j}$. This is the
counter-rotating contribution. In the kinetic regime considered here, we
assume that the noise spectrum is negligible at this sum frequency:
\begin{equation}
\widehat R_j\left(\omega_k+\omega_{k-k_j}\right)
\ll
\widehat R_j\left(\omega_k-\omega_{k-k_j}\right).
\label{rotating-wave-condition}
\end{equation}
It therefore averages out on the slow kinetic time scale and is neglected.
If retained, it would describe a parametric exchange of wave action with the
prescribed random medium rather than conservative mode-to-mode scattering. In other words, the OU process is fast compared with the kinetic evolution but slow enough that its spectrum is negligible at the sum frequency.

Repeating the same second-order calculation for transitions leaving mode $k$
gives the corresponding outgoing contribution. Collecting the incoming and
outgoing transfers yields the master equation
\begin{equation}
\partial_tN_k
=
\sum_j\left[
{\mathcal G}_{k,k_j}N_{k-k_j}
-{\mathcal G}_{k+k_j,k_j}N_k
\right].
\label{general-wave-scattering}
\end{equation}
The first term represents action received by mode $k$ from mode $k-k_j$,
whereas the second represents action transferred from mode $k$ to mode
$k+k_j$. Summing over the complete Fourier lattice gives
\begin{align}
\frac{\dif}{\dif t}\sum_kN_k
&=
\sum_{k,j}{\mathcal G}_{k,k_j}N_{k-k_j}
-\sum_{k,j}{\mathcal G}_{k+k_j,k_j}N_k
\nonumber\\
&=0,
\end{align}
after the change of summation variable $k\mapsto k+k_j$ in the first term.

In the long-wave asymptotic regime $|k|/|k_j|\to0$, the evenness of the
dispersion relation gives
\begin{equation}
\Omega_{k+k_j,-k_j}
\approx
-\Omega_{k,-k_j}.
\end{equation}
Since the stationary covariance is even, the frequency-dependent part of the
transition rate is consequently the same for the above forward and
backward transitions. We additionally assume that the complete coupling
coefficient has the same leading symmetry, so that
\begin{equation}
{\mathcal G}_{k+k_j,k_j}
=
{\mathcal G}_{k,k_j}+o(1).
\label{translation-invariant-kernel}
\end{equation}
To leading order, \eqref{general-wave-scattering} therefore reduces to
\begin{equation}
\partial_tN_k
=
\sum_j{\mathcal G}_{k,k_j}
\left(N_{k-k_j}-N_k\right).
\label{wave-scattering-RM}
\end{equation}
This equation describes conservative mode-to-mode scattering by a prescribed
stationary random medium. The kernel ${\mathcal G}_{k,k_j}$ sets the
characteristic time scale for the redistribution of resolved wave action.
The unresolved modes act here as stochastic transport fluctuations rather
than as dynamically evolving weakly nonlinear wave modes. This mechanism is
therefore distinct from classical three- or four-wave resonant interactions.

 \subsection{Diffusive continuous limit}
 We now consider the continuous wavenumber limit of the master equation \eqref{wave-scattering-RM}. Assuming that the unresolved scattering wavevectors form a sufficiently
dense set and that the discrete scattering kernel admits a regular
continuum representation, the discrete sum may be replaced by an
integral, yielding
 \begin{equation}
 \partial_t N(k) = \int {\mathcal G}(k,q)[N(k-q) -N(k)] \dif q.
 \label{int-mod-E}
 \end{equation}
 Here $q$ denotes the unresolved scattering wavevector and the continuum scattering kernel ${\mathcal G}(k,q)\ge0$ incorporates the density of unresolved scattering modes. 
 
 When the dispersion relation varies slowly in space and time, the wave action also depends on the physical position $x$.The phase, ${\cal P}(x,t)$, satisfies the Eikonal equation,
 \[
 \partial_t {\cal P} +\omega(x,\nabla {\cal P} ,t) =0
 \] 
where the local wavevector is  \(k=\nabla_{\!x}{\cal P}\). The associated ray equations are 
\[
\dot{x} =\nabla_{\!k}\, \omega  \text{ and } \dot{k} = -\nabla_{\!x}\, \omega.\] 
The corresponding wave action density then reads 
 \begin{equation}
  \partial_t N(x,k,t) + \nabla_{\!k}\, \omega \bcdot \nabla_{\!x} N(x,k,t) - \nabla_{\!x} \omega \bcdot \nabla_{\!k} N(x,k,t) =  \int {\mathcal G}(k,q)[N(x,k-q,t) -N(x,k,t)] \dif q.
 \end{equation}
This coincides with the linear radiative transfer equation derived by \cite{Ryzhik-Papanicolaou-Keller-96}.
Under the reciprocity condition
\begin{equation}
{\mathcal G}(k,q)
=
{\mathcal G}(k-q,-q),
\label{scattering-reciprocity}
\end{equation}
the scattering operator conserves total wave action:
\[
\frac{\dif}{\dif t}\int_{\mathbb R^d}N(k,t)\,\dif k=0.
\]
The following argument concerns the scattering dynamics locally in
physical space, with the slow spatial and temporal variables frozen.
In the long-wave regime, we retain the leading-order
translation-invariant part of the scattering kernel,
\[
{\mathcal G}(\varepsilon\kappa,q)
=
{\mathcal G}_0(q)+o(1).
\]
We assume that this leading-order kernel is symmetric,
\[
{\mathcal G}_0(-q)={\mathcal G}_0(q),
\]
as follows, in particular, under isotropy assumption.

We now interpret the scattering dynamics as a random jump process in
spectral space. Let \(K_t\) denote the process associated with the
leading-order translation-invariant scattering operator
\begin{equation}
({\cal L}f)(k)
=
\int_{\mathbb R^d}
{\mathcal G}_0(q)
\bigl[
f(k-q)-f(k)
\bigr],\dif q.
\label{jump-generator-k}
\end{equation}
Thus, the process undergoes spectral jumps
\[
K_t\longmapsto K_t-q
\]
with rate measure \({\mathcal G}_0(q)\dif q\).

To describe the dynamics in the stretched long-wave spectral
coordinate, we first set
\begin{equation}
k=\varepsilon\kappa
\end{equation}
and define
\begin{equation}
\kappa_t^\varepsilon
:=
\varepsilon^{-1}K_t.
\label{kappa-process}
\end{equation}
At this stage, only the spectral variable is rescaled; the time
variable is unchanged. An original jump
\[
K_t\longmapsto K_t-q,
\]
therefore becomes
\begin{equation}
\kappa_t^\varepsilon
\longmapsto
\frac{K_t-q}{\varepsilon}
=
\kappa_t^\varepsilon-\varepsilon^{-1}q.
\end{equation}
The generator of \(\kappa^\varepsilon\) is consequently
\begin{equation}
({\cal L}_{\kappa}^{\varepsilon}f)(\kappa)
=
\int_{\mathbb R^d}
{\mathcal G}_0(q)
\left[
f\left(\kappa-\varepsilon^{-1}q\right)
-
f(\kappa)
\right]\dif q.
\label{kappa-generator}
\end{equation}
Hence the jumps are of order \(\varepsilon^{-1}\) in the stretched
coordinate \(\kappa\), while their rate remains of order one. The
long-wave spectral rescaling alone therefore does not produce a
diffusive limit: on the original time scale, the jumps of
\(\kappa^\varepsilon\) are not asymptotically small, and a martingale
functional central limit theorem cannot be applied directly.

A Brownian limit requires the accumulation of many centered scattering
events over a longer time scale, together with the corresponding
central-limit normalization. We therefore introduce
\begin{equation}
Y_\tau^\varepsilon
:=
\varepsilon^2
\kappa_{\tau/\varepsilon^2}^\varepsilon.
\label{Y-process}
\end{equation}
Equivalently,
\begin{equation}
Y_\tau^\varepsilon
=
\varepsilon K_{\tau/\varepsilon^2}.
\end{equation}
This two-step construction separates two distinct operations. The
definition
\(\kappa_t^\varepsilon=\varepsilon^{-1}K_t\) implements the stretched
long-wave spectral coordinate, without changing time. The accelerated
time \(t=\tau/\varepsilon^2\) then allows
\(\mathcal O(\varepsilon^{-2})\) scattering events to accumulate,
while the prefactor \(\varepsilon^2\) provides the central-limit
normalization appropriate to the order-\(\varepsilon^{-1}\) jumps of
\(\kappa^\varepsilon\). It is therefore not a second long-long-wave
rescaling.

A jump of \(\kappa^\varepsilon\) of size
\(-\varepsilon^{-1}q\) then produces a jump of
\(Y^\varepsilon\) of size
\begin{equation}
\Delta Y^\varepsilon
=
-\varepsilon q.
\end{equation}
Thus, although the jumps are large in the intermediate coordinate
\(\kappa\), they become asymptotically small in the macroscopic
variable \(Y^\varepsilon\).
The time acceleration by \(\varepsilon^{-2}\) compensates for the
vanishing \(\mathcal{O}(\varepsilon^2)\) variance of each rescaled jump. Thus,
jumps of size \(\mathcal{O}(\varepsilon)\) occur at an effective rate
\(\mathcal{O}(\varepsilon^{-2})\), yielding an order-one quadratic variation
and hence a nontrivial Brownian limit.
As a matter of fact, the generator of \(Y^\varepsilon\) is
\begin{equation}
({\cal L}_{Y}^{\varepsilon}g)(y)
=
\varepsilon^{-2}
\int_{\mathbb R^d}
{\mathcal G}_0(q)
\bigl[
g(y-\varepsilon q)-g(y)
\bigr],\dif q.
\label{Y-generator}
\end{equation}
This describes jumps \(-\varepsilon q\), of magnitude
\(\mathcal O(\varepsilon|q|)\), occurring at rate
\(\varepsilon^{-2}{\mathcal G}_0(q)\dif q\).
The symmetry of the jump measure implies that it is centered,
\begin{equation}
\int_{\mathbb R^d}
q\,{\mathcal G}_0(q)\,\dif q
=0,
\label{centered-jumps}
\end{equation}
and has finite second moment,
\begin{equation}
\int_{\mathbb R^d}
|q|^2{\mathcal G}_0(q)\,\dif q
<\infty.
\label{finite-second-moment}
\end{equation}
For every \(\delta>0\), the jumps of \(Y^\varepsilon\) satisfy the negligible jump condition
\begin{equation}
\varepsilon^{-2}
\int_{{|\varepsilon q|>\delta}}
|\varepsilon q|^2
{\mathcal G}_0(q)\,\dif q
=
\int_{{|q|>\delta/\varepsilon}}
|q|^2{\mathcal G}_0(q)\,\dif q
\xrightarrow[\varepsilon\to0]{}0.
\label{lindeberg-condition}
\end{equation}
Hence the jumps of the rescaled process are asymptotically
negligible, even though the jumps in the intermediate coordinate
\(\kappa\) are of order \(\varepsilon^{-1}\).

Applying Dynkin's formula to the coordinate functions
\(g_i(y)=y_i\), condition \eqref{centered-jumps} gives
\[
{\cal L}_{Y}^{\varepsilon}g_i=0.
\]
Consequently,
\begin{equation}
Y_\tau^{\varepsilon,i}
=
Y_0^{\varepsilon,i}
+
M_\tau^{\varepsilon,i},
\end{equation}
where \(M^\varepsilon\) is a martingale. Its predictable quadratic
covariation is
\begin{equation}
\left\langle
M^{\varepsilon,i},
M^{\varepsilon,j}
\right\rangle_\tau
=
\varepsilon^{-2}
\int_0^\tau
\int_{\mathbb R^d}
(\varepsilon q_i)(\varepsilon q_j)
{\mathcal G}_0(q)\,\dif q\,\dif s
=
\tau
\int_{\mathbb R^d}
q_iq_j{\mathcal G}_0(q)\,\dif q.
\label{quadratic-covariation}
\end{equation}
Assuming additionally that
\(Y_0^\varepsilon\Longrightarrow Y_0\), by the multidimensional martingale functional central
limit theorem, Theorem~7.1.4(b) of
\cite{Ethier-Kurtz-86} (see appendix for a statement of this theorem),
\begin{equation}
Y^\varepsilon
\Longrightarrow
Y_0 + \sqrt{2D}\,W,
\label{Y-brownian-limit}
\end{equation}
in the space \(\mathbb D([0,T];\mathbb R^d)\) of c\`adl\`ag  functions (right continuous with left limits) where
\begin{equation}
D^{ij}
=
\frac12
\int_{\mathbb R^d}
q_iq_j{\mathcal G}_0(q)\,\dif q.
\label{diffusion-tensor}
\end{equation}
To translate the convergence of the jump process into a limit for the
wave-action density, we introduce the rescaled density associated with
\(Y^\varepsilon\),
\begin{equation}
N^\varepsilon(y,\tau)
:=
\varepsilon^{-d}
N\left(\frac{y}{\varepsilon},
       \frac{\tau}{\varepsilon^2}\right).
\label{rescaled-action-density}
\end{equation}
The factor \(\varepsilon^{-d}\) ensures preservation of the total wave
action. The rescaled density satisfies
\begin{equation}
\partial_\tau N^\varepsilon
=
\varepsilon^{-2}
\int_{\mathbb R^d}
{\mathcal G}_0(q)
\left[
N^\varepsilon(y-\varepsilon q,\tau)
-
N^\varepsilon(y,\tau)
\right]\dif q.
\label{rescaled-master-equation}
\end{equation}
Equivalently, for every smooth compactly supported test function
\(\varphi\),
\begin{equation}
\frac{\dif}{\dif\tau}
\int_{\mathbb R^d}\varphi(y)N^\varepsilon(y,\tau)\,\dif y
=
\int_{\mathbb R^d}
N^\varepsilon(y,\tau)
({\cal Q}^{\varepsilon,*}\varphi)(y)\,\dif y,
\end{equation}
where the adjoint ${\cal Q}^{\varepsilon,*}$ acting on test functions (while ${\cal Q}^{\varepsilon}$ acts on density) is 
\begin{equation}
({\cal Q}^{\varepsilon,*}\varphi)(y)
=
\varepsilon^{-2}
\int_{\mathbb R^d}{\mathcal G}_0(q)
\left[
\varphi(y+\varepsilon q)-\varphi(y)
\right]\dif q.
\end{equation}
If the scattering kernel is symmetric,
\({\mathcal G}_0(-q)={\mathcal G}_0(q)\), then
\({\cal Q}^{\varepsilon,*}={\cal L}_{Y}^{\varepsilon}\).
At the level of the wave-action density, the same limiting dynamics is
identified through convergence of the generators on smooth test
functions:
\[
{\cal L}_{Y}^{\varepsilon}\varphi
\longrightarrow
D^{ij}\partial_{ij}\varphi,
\]
where $D$ is given by \eqref{diffusion-tensor}.
Hence, if \(N^\varepsilon\) converges weakly to \(\overline N\), then
\[
\frac{\dif}{\dif\tau}\int\varphi\overline N\,\dif y
=
\int\overline N D^{ij}\partial_{ij}\varphi\,\dif y
=
\int\varphi\,
\partial_i\partial_j(D^{ij}\overline N)\,\dif y.
\]
Relabelling the macroscopic variables
\((y,\tau,\overline N)\) as \((k,t,N)\), the limiting equation becomes
\begin{equation}
\partial_tN
=
\nabla_k\bcdot(D\nabla_kN).
\end{equation}
It describes a random walk in spectral space generated by repeated
scattering events. No corrector is required here: because the jump kernel is
translation invariant and centered, the coordinate process is already
a martingale. The diffusion therefore arises directly from its
quadratic variation.

The present diffusion is obtained in a long-wave, long-time
central-limit regime. The microscopic scattering wavevector \(q\) is
not assumed to be small: it is the rescaled jump \(\varepsilon q\)
that becomes small in the long-wave coordinate, while
\(\mathcal O(\varepsilon^{-2})\) centered scattering events accumulate
over one unit of macroscopic time. This regime should be distinguished
from the small-jump approximation that will be considered afterward. In that case the wavenumber variable is not rescaled, and the spectral displacement
itself is small because the interacting resolved and unresolved
wavevectors are close. The corresponding diffusion is then obtained
from a local expansion of the collision operator.

\subsection{Scaling analysis}
In the deep-water regime, the dispersion relation scales as $\omega\sim |k|^{1/2}$, and the Dirichlet-to-Neuman operator satisfies $G_k[0]\sim |k|$. The noise coupling coefficient scales as $|{\cal C}_{k,q}| \sim |c_q| |k|^{1/2} |q|^{3/2}$. The noise amplitude scales as $c_{k,q}\sim |q|^{-2} \omega^{1/2} = |q|^{-7/4}$, we obtain $|{\cal C}_{k,q}| \sim  |k|^{1/2}|q|^{-1/4}$. The scattering kernel therefore behaves as
\begin{equation}
{\mathcal G}(k,q) = \frac{|{\cal C}_{k,q}|^2\sigma_q^2}{\gamma_q^2 + \Omega^2_{k,-q}} \sim \frac{|k| |q|^{-1/2} \sigma_q^2}{\gamma_q^2 + |q|}.
\end{equation}
The diffusion coefficient in wave-number space then scales as
\begin{equation}
D(k)\sim |k|\int_{|k|}^\infty |q|\frac{|q|^{3/2}\sigma_q^2}{\gamma_q^2 + |q|} \dif |q|, 
\end{equation}
If the integral is k-independent (e.g. dominated away from the lower cutoff), one has $D(k)\sim |k|$. When the lower cutoff controls the integral, additional k-dependence arises, as analyzed below. The stationary energy corresponds to a constant wave action flux $\Pi(k) = - D(k) \partial_k N(k)$ with then:
\[
\partial_k N(k)\sim \frac{\Pi}{D(k)},
\]
To further specify the spectrum, we assume OU process parameters follow power laws $\gamma^2_q\sim |q|^\beta$ and $\sigma^2_q\sim |q|^\alpha$. Our goal is to calibrate the noise statistics such that the energy spectrum follows the classical  Kolmogorov-Zakharov wave action slope for nonlinear waves ($|k|^{-4}$). With these assumptions, in radial cooordinates and dimension 2, the diffusion scales as
\begin{equation}
D(k) \sim |k|\int_{|k|}^\infty \frac{|q|^{5/2+\alpha}}{|q|^{\beta}+ |q|} \dif |q|
\end{equation}
Two asymptotic regimes arise: (i) memory dominated regime with $\gamma_q^2\gg\Omega^2_{k,q}$ which implies 
\begin{equation}
D(k) \sim |k|\int |q|^{5/2+\alpha- \beta} \dif |q|,
\label{int-dif1}
\end{equation}
 and (ii) resonnance dominated $\gamma_q^2\ll\Omega^2_{k,q}$ where 
 \begin{equation}
D(k) \sim |k| \int |q|^{3/2+\alpha} \dif |q|.
\label{int-dif2}
\end{equation}
For the diffusion approximation to be controlled by the lower cutoff  $q_{\rm{min}} \sim k$,  the scattering kernel must decay sufficiently fast for large $q$. This requires $\alpha <-\frac{5}{2}$  in the resonance-dominated regime and  $\alpha-\beta < -7/2$  in the memory-dominated regime, which corresponds to turbulent regimes with negative spectrum slope for the noise. We obtain the scaling laws  $N(k)\sim |k|^{-(7/2+\alpha-\beta)}$ in the memory dominated regime and  $N(k)\sim |k|^{-(5/2+\alpha)}$ in the resonance regime.

Matching the Kolmogorov--Zakharov spectrum $N(k)\sim |k|^{-4}$
would formally require $\alpha=\frac32$ 
in the resonance-dominated regime, and $\alpha-\beta=\frac12$
in the memory-dominated regime.
However these exponents violate the convergence conditions
$\alpha<-\frac52$, $\alpha-\beta<-\frac72$
required for the diffusion approximation to be controlled by the lower cutoff.
Therefore, within the present diffusive closure and scaling
assumptions, the stochastic transport mechanism considered here cannot
reproduce the Kolmogorov--Zakharov spectrum. This suggests that induced
spectral cascades driven by stochastic transport noise belong to a
different universality class from weak-turbulence Kolmogorov--Zakharov
cascades. 

 \subsection{Surface-induced additive forcing}
 We now reintroduce the additive forcing produced by the free surface previously neglected. The amplitude evolution reads
 \begin{equation}
 \partial_t a_k = -i\omega_k a_k +\frac{1}{2} \sum_j{\cal C}_{k,k_j} (a_{k-k_j} - a^*_{-(k-k_j)})W^\tau_j(t) +f_k.
 \end{equation}
 
with forcing
 \begin{equation}
  f_k(t) = \sum_j G_k[0] c_{k_j} W^\tau_j(t) \delta_{k,k_j}.
 \end{equation}
After removing fast oscillations, we obtain
  \begin{equation}
 \partial_t b_k =  \frac{1}{2} \sum_j {\cal C}_{k,k_j} b_{k-k_j}Z_j(t) e^{i \Omega_{k,-k_j}t} - \frac{1}{2} \sum_j {\cal C}_{k,k_j} b^*_{-(k-k_j)}Z_j(t) e^{i \Omega_{k,k_j}t} + \widetilde f_k(t),
 \end{equation}
where $\widetilde f_k(t)= f_k(t) e^{i \omega_k t}$.  The energy evolution satisfies
\[
\partial_t N_k = \Exp(\partial_t b_k b^*_k) + \Exp(b\partial_t b^*_k),
\] 
which we decompose as 
\[
\partial_t N_k ={\cal T}_k + {\cal I}_k,
\]
where ${\cal T}_k$ is the conservative transport contribution and
\begin{equation}
{\cal I}_k =\Exp (\widetilde f_k b^*_k) + \Exp (b_k \widetilde f_k^*)
 \end{equation}
 represents additive energy injection.
 
 Using Duhamel expansion,
 \[
 b_k(t) = b_k(0) + \int_0^t  \widetilde f_k(s)\dif s + {\mathcal O}({\mathcal C}),
 \] 
 and the same weak coupling argument as before, we obtain
 \begin{equation} 
 \Exp \bigl( f_k(t)b^*_k(t)\bigr)\approx \int_0^t \Exp \bigl(f_k(t)f^*_k(s)\bigr) \dif s.
 \end{equation}
 Since $W^\tau$ is an OU process, 
 we find,
 at long time, 
 \begin{equation}
 {\mathcal I}_k =  \sum_j|G_{k_j}[0]|^2 \, \frac{\sigma_{k_j}^2}{\gamma_j^2+\omega_k^2} \Exp |c_{k_j}|^2\delta_{k,k_j}.
 \end{equation}
 The full modal energy equation becomes
 \begin{equation}
 \sum_k \partial_t N_k =  \sum_{k,k_j}{\mathcal G}_{k,k_j} (N_{k-k_j} -N_k) + {\cal I}_k,
 \end{equation}
The transport-noise is conservative, while the additive forcing provides localized energy injection in wave-number space. The total wave action density therefore grows according to
\begin{equation}
\frac{\dif}{\dif t}\sum_kN_k = \sum_k  {\cal I}_k.
\end{equation}
When $G_k[0] \varphi^\tau =0$, the additive forcing vanishes.This corresponds to a harmonic velocity potential trace at the surface (shallow water) or to a fractional Laplace condition of order $1/2$ (deep water). Through Helmoltz  decomposition, the remaining forcing corresponds to a weakly compressible surface velocity, 
\begin{equation}
G_k[0] \varphi^\tau = \alpha \chi^\tau_k. 
\end{equation} 
At the energy level, the additive forcing appears  at order $\alpha^2$,
\begin{equation}
{\cal I}_k = \alpha^2 \frac{\sigma_k^2}{\gamma_k^2+\omega_k^2}\Exp |\chi^\tau_k|^2.
\end{equation} 
This injection is balanced by dissipation due to whitecapping and wave breaking, modeled by a linear damping term, $-\mu_k b_k$, which contributes  $-2\mu_k N_k$ to the wave-action equation. A common parameterization  is 
$\mu_k\sim |k|^p$ with $p\geq 2$, consistent with the scale-selective character of whitecapping at small scales.
The final forced-dissipative modal kinetic equation reads
 \begin{equation}
 \partial_t N_k = \sum_{j}{\mathcal G}_{k,k_j} (N_{k-k_j} -N_k) +  \alpha^2 \frac{\sigma_k^2}{\gamma_k}\Exp |\chi_k|^2 - 2 \mu_k N_k.
 \end{equation}
 The multiplicative transport noise redistributes wave action conservatively among modes, whereas the additive contribution originates from stochastic forcing of the free-surface kinematic condition and injects wave action at the unresolved forcing wavenumbers. The resulting kinetic equation therefore contains both conservative scattering and surface-induced production. In a statistically steady regime, injection and dissipation balance; the forcing fixes the flux amplitude, while dissipation arrests the cascade at small scales. The injection rate depends not only on the noise amplitude but also on the memory parameter: shorter correlation times suppress energy input, whereas longer correlation times enhance stochastic production.
 
In the following we now derive a kinetic equation for the complete linear water-waves system. 

\section{Linear stochastic water-wave system with nonstationary correlation modes noise}

In this second case, we consider the complete linear stochastic
water-wave system in which the noise correlation modes evolve
dynamically. Linearizing equations
\eqref{Dyn-Eta}--\eqref{Dyn-LS-Varphi} and \eqref{varphi-i-sto} about a
quiescent free surface, and retaining terms up to first order in the
wave-slope parameter $\epsilon_\mu$, yields
\begin{align}
 &\partial_t \overline\varphi
   + \eta
   + \nabla\varphi^\tau\bcdot\nabla\overline\varphi=0,
   \label{eq-adim-SNSW}\\
 &\partial_t\eta
   =\left.\partial_z\phi\right|_{z=0}
   +\left.\partial_z\phi^\tau\right|_{z=0},\\
 &\partial_t\varphi_i
   +\epsilon_\mu\nabla\overline\varphi\bcdot\nabla\varphi_i=0.
   \label{SV-CF}
\end{align}
The first two equations coincide with the linear stochastic
water-wave system considered in the fixed-noise case. The third
equation describes the evolution of the spatial noise correlation
functions $\varphi_i$. These modes are transported by the large-scale
flow generated by $\overline\varphi$. This is consistent with the
Kraichnan--Tennekes random-sweeping hypothesis, according to which
small-scale structures are advected by energy-containing large-scale
motions. Unlike in the classical random-sweeping picture, this
transport is not postulated here, but follows from the stochastic
variational formulation. As in the previous section, the scaling
associated with large-scale advection is absorbed into the amplitude
of the stochastic forcing.

Equation \eqref{SV-CF} governs the deformation of the spatial
correlation modes only. We now additionally assume that each
unresolved mode possesses an intrinsic Airy carrier. This carrier is
introduced through its temporal stochastic coefficient and not
through the transported spatial mode itself.
\subsection{Wave-amplitude evolution equation}

We derive the wave-amplitude equation for the linear stochastic
system coupled to the small-scale transport equation \eqref{SV-CF}.
Let $Y(s;t,x)$ denote the backward characteristic ending at $x$ at
time $t$, defined by
\begin{equation}
 \frac{\dif}{\dif s}Y(s;t,x)
 =\epsilon_\mu\nabla\overline\varphi\bigl(Y(s;t,x),s\bigr),
 \qquad
 Y(t;t,x)=x,
 \qquad 0\leq s\leq t.
 \label{backward-characteristic}
\end{equation}
We define the inverse flow map by
\[
 X_t^{-1}(x):=Y(0;t,x).
\]
The solution of \eqref{SV-CF} is then
\[
 \varphi_j(x,t)
 =\varphi_j\bigl(X_t^{-1}(x),0\bigr).
\]
For the initial spatial mode
\[
 \varphi_j(x,0)=c_{k_j}e^{ik_j\bcdot x},
\]
we obtain
\begin{equation}
 \varphi_j(x,t)
 =c_{k_j}e^{ik_j\bcdot X_t^{-1}(x)}.
 \label{transported-spatial-mode}
\end{equation}

To attach an Airy carrier to this transported spatial pattern, let
$W_j^\tau(t)$ denote a zero-mean complex stochastic envelope
and set
\begin{equation}
 Z_j^\tau(t)
 :=e^{-i\omega_{k_j}t}W_j^\tau(t),
 \qquad
 \omega_{k_j}^2=gG_{k_j}[0].
 \label{Airy-carried-noise}
\end{equation}
The corresponding real-valued noise is obtained by adding the
complex-conjugate component:
\begin{equation}
 \varphi^\tau(x,t)
 =\sum_{j\in\mathcal J_+}
 \left[
 c_{k_j}e^{ik_j\bcdot X_t^{-1}(x)}Z_j^\tau(t)
 +c_{k_j}^*e^{-ik_j\bcdot X_t^{-1}(x)}
       Z_j^{\tau,*}(t)
 \right].
 \label{full-Airy-transported-noise}
\end{equation}
where $\mathcal J_+$ contains one representative of each pair of
opposite wavevectors $\{k_j,-k_j\}$.
Hereafter, the formulae are written for the positive-frequency branch;
the complex-conjugate contribution required by reality is understood.

In the Ornstein--Uhlenbeck approximation used below for the kinetic
closure, definition \eqref{Airy-carried-noise} is equivalently realized
in law by the complex oscillatory OU process
\begin{equation}
 \dif Z_j
 =-\bigl(\gamma_j+i\omega_{k_j}\bigr)Z_j\,\dif t
 +\sigma_j\,\dif B_j^{\mathbb C}(t),
 \label{oscillatory-OU}
\end{equation}
where $B_j^{\mathbb C}$ are independent complex Brownian
motions. In its stationary regime,
\begin{equation}
 \Exp\!\left[Z_j(t)Z_\ell^*(s)\right]
 =\delta_{j\ell}\frac{\sigma_j^2}{2\gamma_j}
 e^{-\gamma_j|t-s|}e^{-i\omega_{k_j}(t-s)}.
 \label{oscillatory-OU-covariance}
\end{equation}
Thus the covariance remains stationary, but its spectrum is a
Lorentzian centered at the Airy frequency $\omega_{k_j}$. The
coefficient $c_{k_j}$ now contains only the time-independent spatial
weight and polarization of the velocity-potential mode. If it is
parameterized by an initial positive-frequency Airy elevation
amplitude $a'_{k_j}(0)$, then the corresponding convention is
\begin{equation}
 c_{k_j}
 =-\frac{i\omega_{k_j}}{G_{k_j}[0]}a'_{k_j}(0).
 \label{noise-Airy-polarization}
\end{equation}
These noise modes coefficients are constant in time, the time dependency of the noise is indicated through the temporal factor $e^{-i\omega_{k_j}t}$ in
$Z_j^\tau$.

The spatial pattern of each noise mode is transported and deformed by
the large-scale flow through $X_t^{-1}$. Its gradient is
\[
 \nabla\!\left(e^{ik_j\bcdot X_t^{-1}(x)}\right)
 =iJ_t^{\mathrm{tr}}(x)k_j
  e^{ik_j\bcdot X_t^{-1}(x)},
 \qquad
 J_t(x):=\nabla_xX_t^{-1}(x).
\]
Consequently, the positive-frequency contribution to the
transport-noise term is
\begin{equation}
 \nabla\varphi^\tau(x,t)\bcdot\nabla\overline\varphi(x,t)
 =\sum_j c_{k_j}Z_j^\tau(t)
 e^{ik_j\bcdot X_t^{-1}(x)}
 i\left(J_t^{\mathrm{tr}}(x)k_j\bcdot
 \nabla\overline\varphi(x,t)\right)
 +\mathrm{c.c.}
 \label{exact-transport-noise}
\end{equation}
Compared with the fixed-correlation-mode system, the phase is
deformed by the flow map and the local noise wavenumber is transformed
by the inverse-flow gradient. In Fourier space, the original
convolution is therefore replaced, before expansion, by a
time-dependent integral transform that mixes modes.

This expression can be made explicit for a smooth deformation close
to the identity. Define
\[
 B(s;t,x):=\nabla_xY(s;t,x),
 \qquad J_t(x)=B(0;t,x).
\]
Differentiating \eqref{backward-characteristic} with respect to $x$
gives
\begin{equation}
 \frac{\dif}{\dif s}B(s;t,x)
 =\epsilon_\mu
 \nabla^2\, \overline\varphi
       \bigl(Y(s;t,x),s\bigr)B(s;t,x),
 \qquad B(t;t,x)=\mathbb I.
 \label{backward-Jacobian}
\end{equation}
Its formal solution is
\begin{equation}
 B(s;t,x)
 ={\cal T}\!{\cal o}\exp\!\left(
   \epsilon_\mu\int_t^s
   \nabla^2\,\overline\varphi
      \bigl(Y(r;t,x),r\bigr)\,\dif r
   \right),
 \label{backward-Jacobian-solution}
\end{equation}
where ${\cal T}\!{\cal o}$ denotes time ordering. To first order in
$\epsilon_\mu$,
\begin{align}
 X_t^{-1}(x)
 &=x-\epsilon_\mu\int_0^t
       \nabla\overline\varphi(x,s)\,\dif s
   +{\cal O}(\epsilon_\mu^2),
 \label{inverse-map-first-order}\\
 J_t(x)
 &=\mathbb I-\epsilon_\mu\int_0^t
       \nabla^2\,\overline\varphi(x,s)\,\dif s
   +{\cal O}(\epsilon_\mu^2).
 \label{inverse-Jacobian-first-order}
\end{align}
In particular,
\begin{equation}
 e^{ik_j\bcdot X_t^{-1}(x)}
 =e^{ik_j\bcdot x}
 \left[
 1-i\epsilon_\mu k_j\bcdot
       \int_0^t\nabla\overline\varphi(x,s)\,\dif s
 \right]
 +{\cal O}(\epsilon_\mu^2).
 \label{noise-phase-first-order}
\end{equation}

Introduce the Lagrangian displacement and its gradient,
\begin{equation}
 \chi(x,t):=\int_0^t\nabla\overline\varphi(x,s)\,\dif s,
 \qquad
 H(x,t):=\nabla\chi(x,t)
 =\int_0^t\nabla^2\,\overline\varphi(x,s)\,\dif s.
 \label{chi-H-definitions}
\end{equation}
Inserting \eqref{inverse-Jacobian-first-order} and
\eqref{noise-phase-first-order} into \eqref{exact-transport-noise}
gives, for the displayed positive-frequency branch,
\begin{multline}
 \nabla\varphi^\tau\bcdot\nabla\overline\varphi
 =\nabla\overline\varphi\bcdot
 \sum_j c_{k_j}Z_j^\tau(t)e^{ik_j\bcdot x}
 \Bigl[
 ik_j
 -\epsilon_\mu
 \Bigl(iH(x,t)k_j-\bigl(k_j\bcdot\chi(x,t)\bigr)k_j\Bigr)
 \Bigr]
 \\
 +\mathrm{c.c.}
 +{\cal O}(\epsilon_\mu^2).
 \label{expanded-transport-noise}
\end{multline}
The leading term is the fixed-correlation-mode contribution, now
carrying the Airy phase through $Z_j^\tau$:
\begin{equation}
 \left(
 \nabla\varphi^\tau\bcdot\nabla\overline\varphi
 \right)^{(0)}
 =\nabla\overline\varphi\bcdot
 \sum_j c_{k_j}Z_j^\tau(t)e^{ik_j\bcdot x}ik_j
 +\mathrm{c.c.}
 \label{leading-transport-noise}
\end{equation}
For a fixed resolved wavenumber $k$, its positive-frequency Fourier
coefficient is given through the convolution in Fourier space
\begin{equation}
 \left(
 \nabla\varphi^\tau\bcdot\nabla\overline\varphi
 \right)^{(0)}_k
 =-\sum_j
 (k-k_j)\bcdot k_j\,c_{k_j}
 \overline\varphi_{k-k_j}(t)Z_j^\tau(t).
 \label{leading-transport-noise-Fourier}
\end{equation}

The first-order correction contains two effects: deformation of the
noise phase by the Lagrangian displacement and deformation of the
noise wavenumber by the inverse-flow Jacobian. To express them in
Fourier space, write
\begin{align*}
 \overline\varphi(x,t)
 &=\sum_p\overline\varphi_p(t)e^{ip\bcdot x},\\
 \nabla\overline\varphi(x,t)
 &=\sum_p ip\,\overline\varphi_p(t)e^{ip\bcdot x},\\
 \nabla^2\,\overline\varphi(x,t)
 &=\sum_p-p\otimes p\,\overline\varphi_p(t)e^{ip\bcdot x},\\
 \chi(x,t)
 &=\sum_q iq
 \left(\int_0^t\overline\varphi_q(s)\,\dif s\right)
 e^{iq\bcdot x}.
\end{align*}
The phase-deformation factor is
\begin{equation}
 (k_j\bcdot\nabla\overline\varphi(x,t))
 (k_j\bcdot\chi(x,t))
 =-\sum_{p,q}
 (k_j\bcdot p)(k_j\bcdot q)
 \overline\varphi_p(t)
 \left(\int_0^t\overline\varphi_q(s)\,\dif s\right)
 e^{i(p+q)\bcdot x}.
 \label{phase-deformation-Fourier}
\end{equation}
After multiplication by $e^{ik_j\bcdot x}$ and projection onto mode
$k$,  its contribution to the transport term is
\begin{equation}
 -\epsilon_\mu\sum_j c_{k_j}Z_j^\tau(t)
 \sum_{p+q=k-k_j}
 (k_j\bcdot p)(k_j\bcdot q)
 \overline\varphi_p(t)
 \int_0^t\overline\varphi_q(s)\,\dif s.
 \label{phase-deformation-contribution}
\end{equation}
Similarly,
\begin{equation}
 \nabla\overline\varphi(x,t)\bcdot H(x,t)k_j
 =\sum_{p,q}
 (p\bcdot q)(q\bcdot k_j)
 \overline\varphi_p(t)
 \left(\int_0^t\overline\varphi_q(s)\,\dif s\right)
 e^{i(p+q)\bcdot x},
 \label{Jacobian-deformation-Fourier}
\end{equation}
and its contribution to the transport term is
\begin{equation}
 -\epsilon_\mu\sum_j c_{k_j}Z_j^\tau(t)
 \sum_{p+q=k-k_j}
 (p\bcdot q)(q\bcdot k_j)
 \overline\varphi_p(t)
 \int_0^t\overline\varphi_q(s)\,\dif s.
 \label{Jacobian-deformation-contribution}
\end{equation}
Define
\begin{equation}
 \Lambda(k_j,p,q)
 :=(k_j\bcdot p)(k_j\bcdot q)
   +(p\bcdot q)(q\bcdot k_j)
 \label{Lambda-vertex}
\end{equation}
and
\begin{equation}
 {\cal S}_{k,k_j}[\overline\varphi](t)
 :=c_{k_j}\sum_{p+q=k-k_j}
 \Lambda(k_j,p,q)\,
 \overline\varphi_p(t)
 \int_0^t\overline\varphi_q(s)\,\dif s.
 \label{S-transport-correction}
\end{equation}
The Fourier coefficient of the complete transport term can then be
written as
\begin{equation}
 \left(
 \nabla\varphi^\tau\bcdot\nabla\overline\varphi
 \right)_k
 =-\sum_j(k-k_j)\bcdot k_j\,c_{k_j}
   \overline\varphi_{k-k_j}Z_j^\tau
 -\epsilon_\mu\sum_jZ_j^\tau
   {\cal S}_{k,k_j}[\overline\varphi]
 +{\cal O}(\epsilon_\mu^2),
 \label{complete-transport-correction}
\end{equation}
with the conjugate noise branch understood.

Using the Airy variable
\begin{equation}
 a_k
 =\frac12\left(
 \eta_k+\frac{iG_k[0]}{\omega_k}\overline\varphi_k
 \right),
 \qquad
 \overline\varphi_p
 =-\frac{i\omega_p}{G_p[0]}
   \left(a_p-a^*_{-p}\right),
 \label{varphi-p-a-p}
\end{equation}
and projecting \eqref{complete-transport-correction} onto $a_k$ gives
\begin{equation}
 \partial_ta_k
 =-i\omega_ka_k
 +\frac12\sum_j{\cal  C}_{k,k_j}
 \left(a_{k-k_j}-a^*_{-(k-k_j)}\right)Z_j^\tau(t)
 -\epsilon_\mu{\cal  R}_k[a](t)+f_k(t)
 +{\cal O}(\epsilon_\mu^2),
 \label{full-amplitude-transported-noise}
\end{equation}
where ${\cal C}_{k,k_j}$ is the same linear coupling coefficient as
in the fixed-correlation-mode case, while
\begin{align}
 {\cal R}_k[a](t)
 &:=\sum_jZ_j^\tau(t){\cal R}_{k,k_j}[a](t),
 \label{R-total-definition}\\
 {\cal R}_{k,k_j}[a](t)
 &:=-\frac{iG_k[0]}{2\omega_k}
 {\cal S}_{k,k_j}[\overline\varphi(a)](t).
 \label{R-mode-definition}
\end{align}
The factor in \eqref{R-mode-definition} results from the projection of
the Bernoulli equation onto the Airy variable; it also fixes the sign
convention in \eqref{full-amplitude-transported-noise}.

Under the same rotating-wave approximation as in the fixed-mode case,
the counter-rotating term proportional to
$a^*_{-(k-k_j)}$ is neglected in the leading linear scattering term.
The resulting effective equation is
\begin{equation}
 \partial_ta_k
 =-i\omega_ka_k
 +\frac12\sum_j{\cal C}_{k,k_j}
 a_{k-k_j}Z_j^\tau(t)
 -\epsilon_\mu{\cal R}_k[a](t)+f_k(t)
 +{\cal O}(\epsilon_\mu^2).
 \label{effective-amplitude-Airy-noise}
\end{equation}
The additive forcing $f_k$ inherits the same Airy-carried temporal
coefficient at the forced noise wavenumbers.

The Airy carrier is particularly transparent in the interaction
representation $b_k(t)=a_k(t)e^{i\omega_kt}$. For example, the first
scattering term in \eqref{effective-amplitude-Airy-noise} contains the
phase
\begin{equation}
 e^{i(\omega_k-\omega_{k-k_j})t}
 e^{-i\omega_{k_j}t}
 =e^{i(\omega_k-\omega_{k-k_j}-\omega_{k_j})t}.
 \label{Airy-shifted-linear-detuning}
\end{equation}
More generally, \eqref{oscillatory-OU-covariance} replaces each bare
frequency mismatch $\Omega$ in the memory integrals by
$\Omega-\omega_{k_j}$. In particular, for the positive-frequency
quartic branch considered in the kinetic closure, the mismatch is
\begin{equation}
 \Omega_{k,p,q,k_j}
 =\omega_p+\omega_q-\omega_k-\omega_{k_j}.
 \label{Airy-shifted-quartic-detuning}
\end{equation}

The term ${\cal R}_k[a]$ is quadratic in the resolved wave amplitudes
and nonlocal in time. It is therefore an explicitly non-Markovian
correction, of first order in $\epsilon_\mu$ and second order in wave
amplitude. It arises from the interaction between the resolved wave
field and the Lagrangian deformation of the Airy-carried stochastic
correlation modes. Equation
\eqref{effective-amplitude-Airy-noise} thus provides the leading-order
approximation of a Lagrangian stochastic wave model in which the
unresolved spatial patterns are transported by the resolved flow while
their temporal coefficients retain the intrinsic Airy oscillations of
the unresolved waves.

\subsection{Kinetic theory for transported Airy-carried noise}

We now derive the contribution of the transported noise modes to the
wave-action equation. As in the fixed-mode case, we introduce the
interaction variables
\begin{equation}
 b_k(t)=a_k(t)e^{i\omega_k t},
 \qquad
 N_k(t)=\Exp\!\left[|b_k(t)|^2\right].
 \label{interaction-variable-transported-noise}
\end{equation}
The temporal coefficient of an unresolved positive-frequency Airy
mode is written
\begin{equation}
 Z_\ell^\tau(t)
 =e^{-i\omega_\ell t}W_\ell^\tau(t),
 \qquad \omega_\ell^2=gG_\ell[0],
 \label{Airy-carrier-kinetic}
\end{equation}
where $W_\ell^\tau$ is a zero-mean circular complex stochastic
envelope. The real noise contains both $Z_\ell^\tau$ and its complex
conjugate. In the OU approximation, the envelope satisfies
\begin{equation}
 \dif W_\ell
 =-\gamma_\ell W_\ell\,\dif t
 +\sigma_\ell\,\dif B_\ell^{\mathbb C}(t),
 \label{OU-envelope-transported-noise}
\end{equation}
and has stationary covariance
\begin{align}
 \Exp\!\left[W_\ell(t)W_m^*(s)\right]
 &=\delta_{\ell m}R_\ell(t-s),
 &
 R_\ell(r)
 &=\frac{\sigma_\ell^2}{2\gamma_\ell}
   e^{-\gamma_\ell|r|},
 \label{OU-envelope-covariance}\\
 \Exp\!\left[W_\ell(t)W_m(s)\right]
 &=0.
 \label{OU-envelope-pseudocovariance}
\end{align}
Equivalently, $Z_\ell$ obeys the oscillatory OU equation
\[
 \dif Z_\ell
 =-\bigl(\gamma_\ell+i\omega_\ell\bigr)Z_\ell\,\dif t
 +\sigma_\ell\,\dif\widetilde B_\ell^{\mathbb C}(t),
\]
and
\begin{equation}
 \Exp\!\left[Z_\ell(t)Z_m^*(s)\right]
 =\delta_{\ell m}R_\ell(t-s)
 e^{-i\omega_\ell(t-s)}.
 \label{Airy-carrier-covariance-kinetic}
\end{equation}
Thus, contractions of the temporal noise must pair $Z_\ell$ with
$Z_\ell^*$; contractions of the form
$\Exp[Z_\ell(t)Z_m(s)]$ vanish for a circular complex envelope.

The linear stochastic-scattering term produces the same master
operator as in the previous section, with the OU spectrum now shifted
by the Airy frequency. For the positive-frequency branch, its rate is
of the form
\begin{equation}
 {\mathcal G}^{\rm A}_{k,k_j}
 =\frac14|{\cal C}_{k,k_j}|^2
 \widehat R_j\!\left(
 \omega_k-\omega_{k-k_j}-\omega_{k_j}
 \right),
 \label{Airy-shifted-scattering-rate}
\end{equation}
where
\begin{equation}
 \widehat R_j(\Omega)
 :=\int_{\mathbb R}R_j(s)e^{i\Omega s}\,\dif s
 =2\,\mathop{\rm Re}\int_0^\infty
 R_j(s)e^{i\Omega s}\,\dif s.
 \label{noise-power-spectrum}
\end{equation}
The conjugate carrier gives the corresponding negative-frequency
branch. With both branches included, the linear scattering operator
can be written in the gain-loss transition-rate form
\begin{equation}
 ({\cal Q}_{\rm sc}N)_k
 =\sum_j\left[
 {\mathcal G}^{\rm A}_{k,k_j}N_{k-k_j}
 -{\mathcal G}^{\rm A}_{k+k_j,k_j}N_k
 \right].
 \label{Airy-scattering-master-operator}
\end{equation}
If the incoming and outgoing rates coincide at leading order 
\(
 {\mathcal G}^{\rm A}_{k,k_j}
 =
 {\mathcal G}^{\rm A}_{k+k_j,k_j}\), the scattering operator reduces
to
\[
 \sum_j{\mathcal G}^{\rm A}_{k,k_j}
 \left(N_{k-k_j}-N_k\right).
\]

We next consider the quadratic correction ${\cal R}_k[a]$ generated
by deformation of the transported noise modes. The Lagrangian
displacement entering this term contains
\begin{equation}
 \int_0^t\overline\varphi_q(s)\,\dif s,
 \qquad
 \overline\varphi_q(s)
 =-\frac{i\omega_q}{G_q[0]}
 a_q(s)+\text{counter-rotating component}.
 \label{displacement-positive-frequency}
\end{equation}
An additional approximation is required before this nonlocal term can
be represented by a time-homogeneous kinetic equation. If $b_q$
varies on a time scale $T_N$ satisfying $\omega_qT_N\gg1$, integration
by parts gives
\begin{equation}
 \int_0^t\overline\varphi_q(s)\,\dif s
 =\frac{a_q(t)}{G_q[0]}
  -\frac{b_q(0)}{G_q[0]}
  +{\cal O}\!\left((\omega_qT_N)^{-1}\right),
 \label{local-displacement-approximation}
\end{equation}
for the positive-frequency component.  We discard the lower-limit term, which retains an explicit dependence
on the arbitrary initial time and therefore generates nonstationary
initial-layer contributions involving \(b_q(0)\). In the kinetic regime
considered here, the observation time is assumed to be long compared
with both the OU correlation time and the wave phase-mixing time, so
that these initial correlations have decayed. We consequently retain
only the local, time-homogeneous response proportional to \(a_q(t)\).

The resonance with two incoming resolved waves is carried by the
complex-conjugate Airy branch of the real noise. Let $\ell$ denote the
physical wavenumber of this unresolved Airy mode. Its conjugate has
the spatial-temporal factor
$e^{-i\ell\bcdot x+i\omega_\ell t}$, and Fourier matching gives
\begin{equation}
 p+q=k+\ell.
 \label{quartic-wavenumber-matching}
\end{equation}
In the notation used in the preceding spatial calculation, this
corresponds to the relabelling $\ell=-k_j$. It is useful to define
\begin{equation}
 \Lambda_\ell^{-}(p,q)
 :=\Lambda(-\ell,p,q)
 = (\ell\bcdot p)(\ell\bcdot q)
   -(p\bcdot q)(q\bcdot\ell).
 \label{conjugate-Lambda}
\end{equation}
Substituting the positive-frequency component
\[
 \overline\varphi_p(t)
 =
 -\frac{i\omega_p}{G_p[0]}a_p(t)
\]
and the local-displacement approximation
\[
 \int_0^t\overline\varphi_q(s)\,\dif s
 \simeq
 \frac{a_q(t)}{G_q[0]}
\]
into the equation for \(a_k\), where the term ${\mathcal R}_k[a]$ contributes with a factor
\(iG_k[0]/(2\omega_k)\) \eqref{R-mode-definition}. The retained positive-frequency contribution to the amplitude
equation for  ${\mathcal R}_k[a]$ is hence 
\begin{equation}
 \left.\partial_ta_k\right|_{\cal R}
 =
 \epsilon_\mu
 \sum_\ell\sum_{p+q=k+\ell}
 c_\ell^*W_\ell^{\tau,*}(t)e^{i\omega_\ell t}
 A_{k\ell}(p,q)\,
 a_p(t)a_q(t),
 \label{quadratic-interaction-a-unsymmetrized}
\end{equation}
where
\begin{equation}
 A_{k\ell}(p,q)
 :=
 \frac{G_k[0]\omega_p}
 {2\omega_kG_p[0]G_q[0]}
 \Lambda_\ell^-(p,q).
 \label{ordered-transport-vertex}
\end{equation}
Here the conjugate Airy carrier has spatial--temporal dependence
\(e^{-i\ell\bcdot x+i\omega_\ell t}\), so that projection onto
wavenumber \(k\) imposes
\[
 p+q=k+\ell.
\]

The convolution sum in
\eqref{quadratic-interaction-a-unsymmetrized} is taken over ordered
pairs. Since \(a_pa_q=a_qa_p\), exchanging \(p\) and \(q\) gives
\begin{align}
 \sum_{p+q=k+\ell}A_{k\ell}(p,q)a_pa_q
 &=
 \frac12\sum_{p+q=k+\ell}
 \left[
 A_{k\ell}(p,q)+A_{k\ell}(q,p)
 \right]a_pa_q \nonumber\\
 &=
 \sum_{p+q=k+\ell}
 V_{k\ell pq}a_pa_q,
 \label{ordered-pair-symmetrization}
\end{align}
where
\begin{equation}
 V_{k\ell pq}
 :=
 \frac{G_k[0]}
 {4\omega_kG_p[0]G_q[0]}
 \left[
 \omega_p\Lambda_\ell^-(p,q)
 +
 \omega_q\Lambda_\ell^-(q,p)
 \right].
 \label{symmetrized-transport-vertex}
\end{equation}
It follows directly that
\[
 V_{k\ell pq}=V_{k\ell qp}.
\]
The additional factor \(1/2\) in
\eqref{symmetrized-transport-vertex} therefore results from
symmetrization of the ordered convolution; the factor
\(G_k[0]/(2\omega_k)\) comes from the Airy projection.

The corresponding contribution to the complete amplitude equation can
thus be written as
\begin{equation}
 \partial_ta_k
 =
 -i\omega_ka_k
 +
 \epsilon_\mu
 \sum_\ell\sum_{p+q=k+\ell}
 c_\ell^*V_{k\ell pq}
 W_\ell^{\tau,*}(t)e^{i\omega_\ell t}
 a_p(t)a_q(t)
 +\cdots ,
 \label{quadratic-interaction-a}
\end{equation}
where the ellipsis denotes the linear scattering, additive forcing,
counter-rotating terms, and initial-layer contributions. As previously, ntroducing $
 b_k(t)=a_k(t)e^{i\omega_kt}$ cancels the  linear Airy contribution in
\eqref{quadratic-interaction-a}  and gives
\begin{equation}
 \left.\partial_tb_k\right|_{\cal R}
 =
 \epsilon_\mu
 \sum_\ell\sum_{p+q=k+\ell}
 c_\ell^*V_{k\ell pq}
 W_\ell^{\tau,*}(t)b_p(t)b_q(t)
 e^{-i\Omega_{k\ell pq}t},
 \label{quadratic-interaction-b}
\end{equation}
where
\begin{equation}
 \Omega_{k\ell pq}
 :=
 \omega_p+\omega_q-\omega_k-\omega_\ell.
 \label{quartic-frequency-mismatch}
\end{equation}

The wavenumber relation \(p+q=k+\ell\), together with the resonant
condition \(\Omega_{k\ell pq}=0\), defines the same
\(2\leftrightarrow2\) kinematic resonance manifold as a four-wave
interaction. Here, however, the mode \(\ell\) is an unresolved
stochastic Airy carrier rather than a resolved Hamiltonian wave mode.
For a finite OU correlation time, the interaction has a finite
frequency width around this resonance manifold.

The wave-action balance associated with \eqref{quadratic-interaction-b}
starts from
\begin{equation}
 \left.\partial_tN_k\right|_{\cal R}
 =2\epsilon_\mu\,\mathop{\rm Re}
 \sum_\ell\sum_{p+q=k+\ell}c_\ell^*V_{k\ell pq}
 \times
 \Exp\!\left[
 W_\ell^{\tau,*}(t)b_p(t)b_q(t)b_k^*(t)
 \right]e^{-i\Omega_{k\ell pq}t}.
 \label{action-balance-quadratic-interaction}
\end{equation}
We assume weak coupling, separation between the OU correlation time
and the kinetic evolution time, and the quasi-Gaussian closure
\begin{align}
 \Exp[b_p b_{p'}^*]&=N_p\delta_{pp'}, \text{ and }
 \Exp[b_p b_{p'}]=0,
 \label{quasi-Gaussian-second-moments}\\
 \Exp[b_p b_q b_{p'}^*b_{q'}^*]
 &=N_pN_q
 \left(\delta_{pp'}\delta_{qq'}
       +\delta_{pq'}\delta_{qp'}\right).
 \label{quasi-Gaussian-fourth-moment}
\end{align}
Under this closure, all third-order amplitude moments vanish. In
particular, the cross term between the linear scattering operator and
\({\cal R}_k\) does not contribute at this order. The leading
contribution of \({\cal R}_k\) is therefore of order
\(\epsilon_\mu^2\). To determine it, one must insert the first Duhamel
correction successively into each of the three factors
\(b_k^*\), \(b_p\), and \(b_q\) appearing in
\eqref{action-balance-quadratic-interaction}. The correction for \(b_k^*\)
follows directly from \eqref{quadratic-interaction-b}, whereas those
for \(b_p\) and \(b_q\) must be extracted from the full amplitude
equation before the rotating-wave truncation.

The direct contraction contributing to mode $k$ is
\begin{equation}
 \left.{\cal Q}^{(+)}_{\cal R}[N]\right|_k
 =\epsilon_\mu^2
 \sum_\ell\sum_{p+q=k+\ell}
 {\cal K}^{(k)}_{k\ell pq}\,N_pN_q,
 \label{quartic-direct-contribution}
\end{equation}
where the two Wick pairings have been included in the nonnegative
rate
\begin{equation}
 {\cal K}^{(k)}_{k\ell pq}
 :=2|c_\ell|^2|V_{k\ell pq}|^2
 \widehat R_\ell(\Omega_{k\ell pq})\geq0.
 \label{quartic-transition-rate}
\end{equation}
For the OU covariance \eqref{OU-envelope-covariance},
\begin{equation}
 \widehat R_\ell(\Omega)
 =\frac{\sigma_\ell^2}
       {\gamma_\ell^2+\Omega^2}.
 \label{OU-envelope-spectrum}
\end{equation}
The imaginary part of the one-sided integral
$\int_0^\infty R_\ell(s)e^{i\Omega s}\,\dif s$ produces a nonlinear
frequency renormalization. It does not contribute to the real
wave-action transfer rate and will not be retained below.

The remaining response terms cannot be obtained by the change
$(p,q)\mapsto(-p,-q)$. Indeed,
\begin{equation}
 \Lambda_\ell^-(-p,-q)
 =\Lambda(-\ell,-p,-q)
 =\Lambda(\ell,p,q),
 \label{Lambda-no-antisymmetry}
\end{equation}
which is not equal, in general, to $-\Lambda_\ell^-(p,q)$. The required
terms instead come from the branches of the full amplitude equation
that are resonant in the equations for $b_p$ and $b_q$. Using
$p+q=k+\ell$ and applying the same local-response approximation and
ordered-pair symmetrization as in
\eqref{symmetrized-transport-vertex}, these equations contain
\begin{align}
 \left.\partial_tb_p\right|_{\cal R}
 &=2\epsilon_\mu c_\ell V^{(p)}_{k\ell pq}
 W_\ell^\tau(t)b_k(t)b_q^*(t)
 e^{i\Omega_{k\ell pq}t},
 \label{response-equation-bp}\\
 \left.\partial_tb_q\right|_{\cal R}
 &=2\epsilon_\mu c_\ell V^{(q)}_{k\ell pq}
 W_\ell^\tau(t)b_k(t)b_p^*(t)
 e^{i\Omega_{k\ell pq}t},
 \label{response-equation-bq}
\end{align}
where
\begin{align}
 V^{(p)}_{k\ell pq}
 &:=\frac{G_p[0]}{4\omega_pG_k[0]G_q[0]}
 \left[
 \omega_k\Lambda(\ell,k,-q)
 -\omega_q\Lambda(\ell,-q,k)
 \right],
 \label{response-vertex-p}\\
 V^{(q)}_{k\ell pq}
 &:=\frac{G_q[0]}{4\omega_qG_k[0]G_p[0]}
 \left[
 \omega_k\Lambda(\ell,k,-p)
 -\omega_p\Lambda(\ell,-p,k)
 \right].
 \label{response-vertex-q}
\end{align}
For the equation of \(b_p\), the product \(b_kb_q^*\) receives two
contributions from the ordered convolution, corresponding to the
placements \((k,-q)\) and \((-q,k)\). Their sum is
\[
 \frac{G_p[0]}
 {2\omega_pG_k[0]G_q[0]}
 \left[
 \omega_k\Lambda(\ell,k,-q)
 -\omega_q\Lambda(\ell,-q,k)
 \right]
 =2V^{(p)}_{k\ell pq}.
\]
Here the factor two restores the two ordered placements of the pair
$(k,-q)$ or $(k,-p)$ after the vertex has been symmetrized. The minus
signs in these vertices arise from the
negative-frequency components of
$\overline\varphi_{-q}$ and $\overline\varphi_{-p}$, respectively.

Inserting \eqref{response-equation-bp} and
\eqref{response-equation-bq} into the mixed moment in
\eqref{action-balance-quadratic-interaction} gives the signed response
coefficients
\begin{align}
 {\cal K}^{(p)}_{k\ell pq}
 &:=-2|c_\ell|^2
 \mathop{\rm Re}\!\left(
 V_{k\ell pq}V^{(p)}_{k\ell pq}
 \right)
 \widehat R_\ell(\Omega_{k\ell pq}),
 \label{response-rate-p}\\
 {\cal K}^{(q)}_{k\ell pq}
 &:=-2|c_\ell|^2
 \mathop{\rm Re}\!\left(
 V_{k\ell pq}V^{(q)}_{k\ell pq}
 \right)
 \widehat R_\ell(\Omega_{k\ell pq}).
 \label{response-rate-q}
\end{align}
The factor two has the same origin as in
\eqref{quartic-transition-rate}: the two ordered placements of the
resolved pair give identical contractions after symmetrization. In
contrast to ${\cal K}^{(k)}$, the coefficients
${\cal K}^{(p)}$ and ${\cal K}^{(q)}$ are not necessarily positive.
They describe the signed response of the two incoming resolved modes
to the same stochastic interaction.

The kinetic operator obtained directly from the three Duhamel
responses is therefore
\begin{equation}
 ({\cal Q}_{\cal R}N)_k
 =\epsilon_\mu^2
 \sum_\ell\sum_{p+q=k+\ell}
 \left[
 {\cal K}^{(k)}_{k\ell pq}N_pN_q
 -{\cal K}^{(p)}_{k\ell pq}N_kN_q
 -{\cal K}^{(q)}_{k\ell pq}N_kN_p
 \right].
 \label{bath-mediated-collision-operator}
\end{equation}
This is a bath-mediated quadratic response operator. In general, it
cannot be written in a single-kernel gain--loss form, since no
equality between its three kernels has been imposed. If, for each admissible quartet \((k,\ell,p,q)\), these kernels satisfy
the quartet-wise common-rate condition
\begin{equation}
 {\cal K}^{(p)}_{k\ell pq}
 =
 {\cal K}^{(q)}_{k\ell pq}
 =
 {\cal K}^{(k)}_{k\ell pq},
 \label{common-response-rate-condition}
\end{equation}
then \eqref{bath-mediated-collision-operator} reduces to the quadratic
gain--loss bracket
\begin{equation}
 ({\cal Q}_{\cal R}N)_k
 =
 \epsilon_\mu^2
 \sum_\ell\sum_{p+q=k+\ell}
 {\cal K}^{(k)}_{k\ell pq}
 \left[
 N_pN_q-N_k(N_p+N_q)
 \right].
 \label{reciprocal-bath-mediated-collision-operator}
\end{equation}
Condition \eqref{common-response-rate-condition} expresses equality
of the elementary production and depletion rates associated with each
interaction quartet. It does not imply that the corresponding action
transfers balance, since they are weighted by the generally different
products \(N_pN_q\), \(N_kN_q\), and \(N_kN_p\).

A sufficient, but stronger and phase-convention-dependent, condition
at the level of the vertices is
\begin{equation}
 V^{(p)}_{k\ell pq}
 =
 V^{(q)}_{k\ell pq}
 =
 -V_{k\ell pq}^{*}.
 \label{reciprocal-adjoint-vertex-condition}
\end{equation}
Relations of this type arise in a closed, canonically normalized
four-wave Hamiltonian system from the reality and permutation
symmetries of a single quartic Hamiltonian. They have not been
established for the present bath-mediated stochastic interaction and
are not assumed below. No factor \(N_\ell\) appears in either form,
because the variance of the unresolved carrier is already contained
in \( |c_\ell|^2\widehat R_\ell \); introducing an additional factor
\(N_\ell\) would count the unresolved spectrum twice.

Combining the linear scattering, the transported-mode correction, the
additive injection ${\cal I}_k$, and the dissipation gives
\begin{multline}
 \partial_tN_k
 =\sum_j\left[
 {\mathcal G}^{\rm A}_{k,k_j}N_{k-k_j}
 -{\mathcal G}^{\rm A}_{k+k_j,k_j}N_k
 \right]
 \\
 +\epsilon_\mu^2
 \sum_\ell\sum_{p+q=k+\ell}
 \left[
 {\cal K}^{(k)}_{k\ell pq}N_pN_q
 -{\cal K}^{(p)}_{k\ell pq}N_kN_q
 -{\cal K}^{(q)}_{k\ell pq}N_kN_p
 \right]
 +{\cal I}_k-2\mu_kN_k.
 \label{kinetic-equ-sto}
\end{multline}
With the normalization used in the preceding additive-forcing
calculation, one may set
\[
 {\cal I}_k
 =\alpha^2\frac{\sigma_k^2}{\gamma_k}|\chi_k|^2.
\]

For the continuum description, we replace the sums by integrals. The
wavenumber constraint is imposed by a Dirac distribution:
\begin{multline}
 \partial_tN_k
 =({\cal Q}_{\rm sc}N)_k
 +\epsilon_\mu^2
 \int_{\mathbb R^{3d}}
 \left[
 {\cal K}^{(k)}_{k\ell pq}N_pN_q
 -{\cal K}^{(p)}_{k\ell pq}N_kN_q
 -{\cal K}^{(q)}_{k\ell pq}N_kN_p
 \right]
 \\
 \times\delta^{(d)}(p+q-k-\ell)
 \,\dif p\,\dif q\,\dif\ell
 +{\cal I}_k-2\mu_kN_k.
 \label{kinetic-equ-continuous-colored}
\end{multline}
For an OU envelope with fixed stationary variance
$D_\ell=\sigma_\ell^2/(2\gamma_\ell)$,
\begin{equation}
 \widehat R_\ell(\Omega)
 =\frac{2D_\ell\gamma_\ell}
       {\gamma_\ell^2+\Omega^2}
 \xrightarrow[\gamma_\ell\to0]{}
 2\pi D_\ell\delta(\Omega).
 \label{narrow-band-resonance-limit}
\end{equation}
This is the long-correlation, narrow-band limit of the Airy carrier,
not the white-noise decorrelation limit. In this limit,
\begin{multline}
 \left.({\cal Q}_{\cal R}N)_k\right|_{\rm res}
 =4\pi\epsilon_\mu^2
 \int_{\mathbb R^{3d}}
 D_\ell|c_\ell|^2
 \Bigl[ |V_{k\ell pq}|^2N_pN_q
 + \mathop{\rm Re} \bigl(
 V_{k\ell pq}V^{(p)}_{k\ell pq}
 \bigr)N_kN_q
+\mathop{\rm Re} \bigl(
 V_{k\ell pq}V^{(q)}_{k\ell pq}
 \bigr)N_kN_p
 \Bigr]
 \\
 \times\delta^{(d)}(p+q-k-\ell)
 \delta(\omega_p+\omega_q-\omega_k-\omega_\ell)
 \,\dif p\,\dif q\,\dif\ell.
 \label{kinetic-equ-resonant-limit}
\end{multline}
By contrast, the white-noise limit is obtained by letting
\(\gamma_\ell\to\infty\) while keeping the covariance intensity
\(Q_\ell\) fixed. The corresponding OU covariance is \(
 R_\ell^\gamma(s)
 =
 \frac{Q_\ell\gamma_\ell}{2}
 e^{-\gamma_\ell|s|},
\)
for which
\[
 R_\ell^\gamma
 \xrightarrow[\gamma_\ell\to\infty]{}
 Q_\ell\delta
\text{ and }
 \widehat R_\ell^\gamma(\Omega)
 =
 Q_\ell\frac{\gamma_\ell^2}
 {\gamma_\ell^2+\Omega^2}
 \longrightarrow Q_\ell
\]
for every fixed \(\Omega\). The limiting temporal spectrum is therefore
flat and does not select the exact resonance
\(\Omega_{k\ell pq}=0\). The Dirac distribution in frequency arises
instead in the opposite, long-correlation limit
\(\gamma_\ell\to0\) with fixed stationary variance. For finite correlation time, the exact resonance is replaced
by a Lorentzian resonance broadening of characteristic frequency width
\(\gamma_\ell\).

The operator \eqref{bath-mediated-collision-operator} shares the
\(2\leftrightarrow2\) quartet geometry \(k+\ell=p+q\) of a
Hasselmann--Zakharov interaction, but its statistical structure
is different. For prescribed carrier statistics, it is quadratic
in the resolved action: the two-time covariance of the unresolved
Airy carrier is already incorporated into the three response
kernels \({\cal K}^{(k)}\), \({\cal K}^{(p)}\), and
\({\cal K}^{(q)}\). These kernels need not satisfy the permutation
identities of a closed Hamiltonian four-wave system. The resolved
dynamics is therefore treated as an open system, whose action
conservation must be checked from the response kernels rather
than inferred from the quartet geometry alone. The relative
signs and weights of the response contributions follow from
the calculated vertices and covariance contractions.

The mechanism may consequently be interpreted as stochastic
wave--wave scattering mediated by unresolved velocity fluctuations
carrying an Airy phase, whose spatial patterns are swept and
deformed by the resolved flow. This connects the construction
with the Kraichnan--Tennekes sweeping picture
\cite{kraichnan1961dynamics,tennekes1975sweeping}
through the advection and temporal decorrelation of unresolved
patterns. The scattering mechanism additionally involves their
deformation and coupling to the resolved waves, and remains
distinct from the intrinsic Hamiltonian four-wave interaction
of classical weak-turbulence theory
\cite{hasselmann1962nonlinear,hasselmann1963nonlinear,
zakharov1966turbulence,zakharov1967weak,zakharov1968stability}.

\paragraph{Remark:}{\bf Relation to four-wave kinetics and higher-order closure}
\label{sec-higher-order-closure}

The distinction can be made explicit by considering the classical
Hasselmann--Zakharov cubic spectral bracket:
\begin{equation}
\begin{aligned}
{\cal B}^{\mathrm{HZ}}_{k\ell pq}[N]
&=
N_pN_q(N_k+N_\ell)
-N_kN_\ell(N_p+N_q)
\\
&=
N_\ell\left[N_pN_q-N_k(N_p+N_q)\right]
+N_kN_pN_q.
\end{aligned}
\label{mixed-Hasselmann-bracket}
\end{equation}
When \(N_\ell\) is associated with the unresolved wave
component, the part proportional to \(N_\ell\) has the same
quadratic spectral bracket as the stochastic operator under
the common-response-rate condition. The separate contribution
\(N_kN_pN_q\), which contains no explicit factor of the unresolved
spectrum, is absent from the retained stochastic closure.

Under the assumed common scaling, with normalized spectra and
unresolved covariance held \(O(1)\), the stochastic transfer is
\(O(\epsilon^2)\), whereas the classical Hasselmann--Zakharov
transfer enters at \(O(\epsilon^4)\). The present model thus
describes the leading stochastic transfer without including
the complete higher-order wave--wave transfer.

To account for this additional transfer, the collision operator
may be extended schematically as
\begin{equation}
\left.\partial_t N\right|_{\mathrm{coll}}
=
\epsilon^2{\cal Q}^{(2)}[N;C^\sigma]
+
\epsilon^4{\cal Q}^{(4)}[N;C^\sigma]
+\cdots,
\label{collision-order-hierarchy}
\end{equation}
where \(C^\sigma\) denotes the unresolved covariance and
\({\cal Q}^{(2)}\) is the leading stochastic operator with
its explicit \(\epsilon^2\) prefactor factored out.
The fourth-order contribution can either be derived by extending
the dynamical expansion and statistical closure consistently
to that order, or modelled through a Hasselmann--Zakharov-type
operator with the full cubic bracket
\eqref{mixed-Hasselmann-bracket} and a quartet-symmetric
interaction kernel. The explicit route includes higher-order
corrections to the inverse flow map and its Jacobian, together
with all other contributions required at the same order.
In the modelling approach, the Hamiltonian four-wave structure
provides an additional closure for the higher-order transfer;
it is not claimed to follow directly from the leading stochastic
operator.

\subsection{Asymptotic scaling}

We now examine the scaling consequences of the kinetic equation
\eqref{kinetic-equ-sto}, or of its continuum counterpart
\eqref{kinetic-equ-continuous-colored}. Two contributions must be
distinguished: the linear scattering operator and the quadratic
bath-mediated interaction generated by deformation of the transported
correlation modes.

For the scattering contribution, we assume that the incoming and
outgoing transition rates are asymptotically equal:
\begin{equation}
 {\mathcal G}^{\rm A}_{k+k_j,k_j}
 =
 {\mathcal G}^{\rm A}_{k,k_j}
 +o(1).
 \label{asymptotically-balanced-Airy-rates}
\end{equation}
The conservative scattering operator
\[
 ({\cal Q}_{\rm sc}N)_k
 =
 \sum_j
 \left[
 {\mathcal G}^{\rm A}_{k,k_j}N_{k-k_j}
 -
 {\mathcal G}^{\rm A}_{k+k_j,k_j}N_k
 \right]
\]
then reduces at leading order to
\begin{equation}
 ({\cal Q}_{\rm sc}N)_k
 =
 \sum_j{\mathcal G}^{\rm A}_{k,k_j}
 \left(N_{k-k_j}-N_k\right)
 +o(1).
 \label{balanced-rate-scattering-operator}
\end{equation}
We refer to \eqref{asymptotically-balanced-Airy-rates} as
translation invariance of the rates at leading order. It should not
be confused with the stronger condition
\({\mathcal G}^{\rm A}(k,q)={\mathcal G}_0(q)\): the common
leading-order rate in \eqref{balanced-rate-scattering-operator} may
retain a slow dependence on the resolved wavenumber \(k\).

\paragraph{Diffusive scattering regime.}
In the continuum limit,
\eqref{balanced-rate-scattering-operator} becomes
\begin{equation}
 ({\cal Q}_{\rm sc}N)(k)
 =
 \int_{\mathbb R^d}
 {\mathcal G}^{\rm A}(k,q)
 \bigl[N(k-q)-N(k)\bigr]\,\dif q .
 \label{balanced-continuum-scattering}
\end{equation}
We consider a scale-local finite-variance regime. The rescaled
scattering variable may satisfy
\[
 |q|\sim{\mathcal k},\qquad {\mathcal k}=|k|,
\]
while the actual spectral displacement is
\(\delta k=\varepsilon q\), with \(\varepsilon\ll1\). Thus
\(|\delta k|\ll{\mathcal k}\), even though the second moment of the
rescaled jump distribution is dominated by modes with
\(|q|\sim{\mathcal k}\). Jumps of size \(\varepsilon q\) are observed
on the accelerated time scale \(\varepsilon^{-2}\). This is the
small-jump, high-rate scaling needed for a local diffusion
approximation. It differs from the preceding long-wave scaling, in
which the unresolved scattering modes appear as large jumps in the
stretched spectral coordinate.

We assume that the kernel has a finite second moment,
\begin{equation}
 \int_{\mathbb R^d}|q|^2{\cal A}(k,q)\,\dif q<\infty .
 \label{variable-kernel-finite-moment}
\end{equation}
For example, a tail
\({\cal A}(k,q)=O(|q|^{-\alpha})\), with
\(\alpha>d+2\), is sufficient. The finite second moment is needed for
the Brownian limit, whereas the small-jump and accelerated-time
scaling makes the resulting diffusion approximation local.

\paragraph{Reciprocity and Taylor expansion.}
When the rate depends on \(k\), symmetry in \(q\) at a fixed
wavenumber is not by itself sufficient to preserve total action at
the next order. Let \(W_\varepsilon(k,k')\) denote the transition rate
between two spectral states and impose reciprocity,
\[
 W_\varepsilon(k,k')=W_\varepsilon(k',k).
\]
Writing \(k'=k-\varepsilon q\), this symmetric edge rate can be
parametrized by its midpoint:
\[
 W_\varepsilon(k,k-\varepsilon q)
 =
 {\cal A}\left(k-\frac{\varepsilon q}{2},q\right),
 \qquad
 {\cal A}(k,-q)={\cal A}(k,q).
\]
The midpoint is therefore not a numerical quadrature rule; it simply
labels the reciprocal transition by the midpoint of its two end
states. The conservative small-jump operator reads
\begin{equation}
 ({\cal Q}_{\rm sc}^{\varepsilon}N)(k)
 =
 \varepsilon^{-2}
 \int_{\mathbb R^d}
 {\cal A}\left(k-\frac{\varepsilon q}{2},q\right)
 \bigl[N(k-\varepsilon q)-N(k)\bigr]\,\dif q ,
 \label{reciprocal-small-jump-operator}
\end{equation}
where
\({\cal A}(k,q)={\mathcal G}^{\rm A}(k,q)+o(1)\).

The Taylor expansions
\begin{align}
 N(k-\varepsilon q)-N(k)
 &=
 -\varepsilon q_i\partial_iN
 +\frac{\varepsilon^2}{2}
 q_iq_j\partial_{ij}N
 +o(\varepsilon^2|q|^2),\\
 {\cal A}\left(k-\frac{\varepsilon q}{2},q\right)
 &=
 {\cal A}(k,q)
 -\frac{\varepsilon}{2}
 q_i\partial_i{\cal A}(k,q)
 +o(\varepsilon|q|)
 \label{small-jump-Taylor-expansions}
\end{align}
make the diffusion limit explicit. Evenness of \({\cal A}\) removes
the order-\(\varepsilon^{-1}\) contribution. The remaining terms give
\begin{equation}
 {\cal Q}_{\rm sc}^{\varepsilon}N
 \longrightarrow
 D^{ij}(k)\partial_{ij}N
 +\partial_iD^{ij}(k)\partial_jN
 =
 \partial_i
 \left(D^{ij}(k)\partial_jN\right),
 \qquad
 D^{ij}(k)
 =
 \frac12\int_{\mathbb R^d}
 q_iq_j{\cal A}(k,q)\,\dif q .
 \label{variable-scattering-diffusion}
\end{equation}
Here symmetry gives \(D^{ij}=D^{ji}\). If the rate were frozen at
\(k\) before expanding, the term
\(\partial_iD^{ij}\partial_jN\) would be absent and the result would
be the nondivergence-form operator
\(D^{ij}\partial_{ij}N\). The reciprocal edge-rate representation is
needed only to retain conservative divergence form for a
variable-coefficient tensor; it makes no difference when \(D\) is
constant.

The leading-order equality of the incoming and outgoing rates does
not force \(D\) to be constant. In \(d\) dimensions,
\[
 [\,{\mathcal G}^{\rm A}\,]
 =
 (\hbox{wavenumber})^{-d}(\hbox{time})^{-1},
 \qquad
 [\,D\,]
 =
 (\hbox{wavenumber})^2(\hbox{time})^{-1}.
\]
In two dimensions,
\([{\mathcal G}^{\rm A}]=L^2T^{-1}\), consistently with the
dimension of \(\sigma_u^2\tau_{\rm eff}\).

\paragraph{Scale-local estimate and transfer rate.}
We now make the physical scaling of the diffusion tensor explicit.
Assume in two horizontal dimensions that the kernel is concentrated
near its lower characteristic scale \(|q|\sim{\mathcal k}\) and may
be represented as
\begin{equation}
 {\cal A}(k,q)
 \sim
 \sigma_u^2({\mathcal k})
 \tau_{\rm eff}({\mathcal k})
 \Phi\left(\frac{|q|}{\mathcal k}\right),
 \label{scale-local-kernel-shape}
\end{equation}
where \(\Phi(\xi)\) is concentrated for \(\xi=O(1)\) and decays
sufficiently rapidly as \(\xi\to\infty\). For each unresolved mode \(k_j\),  the normalized one-sided
covariance integral defines the resonance-weighted correlation time
\begin{equation}
\tau_{{\rm eff},j}(\Omega_{k,k_j})
:=
\mathop{\rm Re}
\int_0^\infty
\frac{R_q(s)}{R_q(0)}
e^{i\Omega_{k,k_j}s}\,\dif s
=
\frac{\gamma_j}
     {\gamma_j^2+\Omega_{k,j}^2}
=
\frac{\gamma^{-1}_{j}}
     {1+(\Omega_{k,j}\gamma^{-1}_{j})^2},
\label{effective-Airy-correlation-time}
\end{equation}
where
\[
\Omega_{k,j}
=
\omega_k-\omega_{k-q}-\omega_q.
\]
In the scaling estimates,
\(\tau_{\rm eff}(k)\) denotes a representative value of
\(\tau_{{\rm eff},j}(\Omega_{k,j})\) for the dominant scattering
wavevectors \(k_j\sim k\). Near resonance,
\(|\Omega_{k,j}|\gamma^{-1}_{j}\ll1\), one has
\(\tau_{\rm eff}(k)\simeq\tau_c(k)\).

If the scattering modes
satisfy \(|q|\gtrsim{\mathcal k}\), the total rate factor is
\begin{align}
 \lambda({\mathcal k})
 &:=
 \int_{\mathbb R^2}{\cal A}(k,q)\,\dif q \nonumber\\
 &\sim
 2\pi\sigma_u^2({\mathcal k})
 \tau_{\rm eff}({\mathcal k})
 {\mathcal k}^2
 \int_1^\infty \xi\Phi(\xi)\,\dif\xi
 \sim
 \sigma_u^2({\mathcal k})
 \tau_{\rm eff}({\mathcal k})
 {\mathcal k}^2 .
 \label{scale-local-total-rate}
\end{align}
Keeping the decaying shape \(\Phi\) in this estimate is essential;
replacing it by a constant inside an integral extending to infinity
would give a divergent expression.

Define the conditional second moment of the rescaled jump by
\begin{equation}
 \left\langle q_iq_j\right\rangle_k
 :=
 \frac{
 \displaystyle\int_{\mathbb R^2}
 q_iq_j{\cal A}(k,q)\,\dif q
 }{
 \displaystyle\int_{\mathbb R^2}
 {\cal A}(k,q)\,\dif q
 }.
 \label{conditional-jump-second-moment}
\end{equation}
Isotropy and dominance by \(|q|\sim{\mathcal k}\) give
\begin{equation}
 \left\langle q_iq_j\right\rangle_k
 \sim
 {\mathcal k}^2\delta_{ij}.
 \label{scale-local-conditional-variance}
\end{equation}
Consequently,
\begin{equation}
 D^{ij}(k)
 =
 \frac12\lambda({\mathcal k})
 \left\langle q_iq_j\right\rangle_k
 \sim
 {\mathcal k}^4
 \sigma_u^2({\mathcal k})
 \tau_{\rm eff}({\mathcal k})
 \delta_{ij}.
 \label{scale-local-diffusion-tensor}
\end{equation}
The time required to diffuse across a spectral distance of order
\({\mathcal k}\) is therefore
\begin{equation}
 \tau_{\rm sc}^{-1}({\mathcal k})
\sim
 \frac{D({\mathcal k})}{{\mathcal k}^2}
 \sim
 {\mathcal k}^2\sigma_u^2({\mathcal k})
 \tau_{\rm eff}({\mathcal k}) .
 \label{scale-local-variable-D}
\end{equation}
The transfer-rate estimate in
\eqref{scale-local-variable-D} will be compared below with the
Hasselmann--Zakharov quartic transfer rate. It is compatible with
\eqref{asymptotically-balanced-Airy-rates}, because equality of the
incoming and outgoing rates does not eliminate their common slow
dependence on \(k\).

\paragraph{Radial diffusion and constant action flux.}
For an isotropic kernel,
\(D^{ij}(k)=D({\mathcal k})\delta_{ij}\). The modal isotropic action
density \(N(k,t)=N({\mathcal k},t)\) satisfies, in two horizontal
dimensions,
\begin{equation}
 \partial_tN
 =
 \frac{1}{\mathcal k}
 \partial_{\mathcal k}
 \left[
 {\mathcal k}D({\mathcal k})
 \partial_{\mathcal k}N
 \right],
 \label{radial-variable-diffusion}
\end{equation}
and the shell-integrated action flux is
\begin{equation}
 \Pi_N({\mathcal k})
 =
 -2\pi{\mathcal k}D({\mathcal k})
 \partial_{\mathcal k}N .
 \label{radial-action-flux-variable-D}
\end{equation}
Here \(2\pi{\mathcal k}N({\mathcal k})\) is the shell-integrated
action density. A nonzero constant shell flux satisfies
\[
 -2\pi{\mathcal k}D({\mathcal k})
 \partial_{\mathcal k}N
 =
 \Pi_N^0.
\]
If \(D({\mathcal k})\sim{\mathcal k}^m\), a nonzero constant radial
flux gives
\begin{equation}
 N({\mathcal k})\sim{\mathcal k}^{-m},
 \qquad m\neq0.
 \label{variable-D-constant-flux-spectrum}
\end{equation}
The limiting case \(m=0\), corresponding to the stronger assumption
\({\mathcal G}^{\rm A}(k,q)={\mathcal G}_0(q)\), gives a logarithmic
profile.

 If
\begin{equation}
 \sigma_u^2({\mathcal k})\sim{\mathcal k}^{-2r},
 \qquad
 \gamma_{\mathcal k}\sim{\mathcal k}^{\beta},
 \label{velocity-memory-power-laws}
\end{equation}
then, near resonance,
\begin{align}
 D({\mathcal k})
 &\sim{\mathcal k}^{4-2r-\beta},
 \label{scale-local-D-power-law}\\
 \tau_{\rm sc}^{-1}({\mathcal k})
 &\sim{\mathcal k}^{2-2r-\beta},
 \label{scale-local-transfer-rate-power-law}\\
 N({\mathcal k})
 &\sim{\mathcal k}^{-(4-2r-\beta)},\\
 E({\mathcal k})=\omega_{\mathcal k}N({\mathcal k})
 &\sim{\mathcal k}^{-7/2+2r+\beta}
 \qquad\hbox{in deep water}.
 \label{scale-local-energy-power-law}
\end{align}
Accordingly, the exponent \(N\sim{\mathcal k}^{-23/6}\) is recovered
only if
\begin{equation}
 2r+\beta=\frac16.
 \label{condition-for-23-over-6}
\end{equation}
For \(r=1/3\), this requires \(\beta=-1/2\), i.e. a correlation time
that grows as \({\mathcal k}^{1/2}\). It does not follow from choosing
\(\gamma_{\mathcal k}\sim\omega_{\mathcal k}\sim{\mathcal k}^{1/2}\),
which instead corresponds to \(\beta=1/2\). The latter choice gives
\[
 \tau_{\rm sc}^{-1}\sim{\mathcal k}^{5/6},
 \qquad
 N({\mathcal k})\sim{\mathcal k}^{-17/6},
 \qquad
 E({\mathcal k})\sim{\mathcal k}^{-7/3}.
\]

Under the conditional choice \(\beta=-1/2\), the value \(r=1/4\)
instead gives
\[
 N({\mathcal k})\sim{\mathcal k}^{-4},
 \qquad
 E({\mathcal k})\sim{\mathcal k}^{-7/2}.
\]
These exponents coincide with familiar Kolmogorov--Zakharov
exponents only at the level of their power laws. Here they arise from
a forced constant-action-flux solution of the stochastic diffusion
operator, rather than from a conservative resonant four-wave
collision integral.

For completeness, multiplying the shell action balance by the
deep-water frequency gives
\begin{equation}
 \partial_t{\mathcal E}_{\rm sh}
 +\partial_{\mathcal k}
 \left(\omega_{\mathcal k}\Pi_N\right)
 =
 \omega_{\mathcal k}'\Pi_N,
 \qquad
 {\mathcal E}_{\rm sh}
 :=
 2\pi{\mathcal k}\omega_{\mathcal k}N .
 \label{diffusive-shell-energy-budget}
\end{equation}
Thus \(\omega_{\mathcal k}\Pi_N\) is not a constant conserved energy
flux: the right-hand side represents the failure of the scattering
diffusion to conserve resolved wave energy. In particular, a constant
inverse action flux does not by itself establish a constant-flux
direct energy cascade.

Finally, subject to decay at infinity or no-flux boundary conditions,
\begin{equation}
 \frac{\dif}{\dif t}
 \left(\frac12\int_{\mathbb R^d}N^2\,\dif k\right)
 =
 -\int_{\mathbb R^d}
 \nabla_{\!k}N\bcdot D(k)\nabla_{\!k}N\,\dif k
 \leq0.
 \label{variable-D-gradient-flow}
\end{equation}
This decay identity describes homogenization in a closed system and
must not be confused with a forced constant-flux stationary state.
The diffusion coefficient may depend parametrically, or through a
closure, on the transported small-scale variance; positivity of
\(D\) is sufficient for the decay identity.

\paragraph{Scaling of the transported correlation modes.}
We next consider the scalar transport equation
\begin{equation}
 \partial_t\varphi_j
 +\epsilon_\mu U\bcdot\nabla\varphi_j
 =0,
 \qquad U=\nabla\overline\varphi .
 \label{transported-mode-scaling-equation}
\end{equation}
The value of \(\varphi_j\) is conserved along a characteristic. Its
spatial \(L^2\) norm, however, obeys
\begin{equation}
 \frac{\dif}{\dif t}
 \int|\varphi_j|^2\,\dif x
 =
 \epsilon_\mu
 \int(\nabla\bcdot U)|\varphi_j|^2\,\dif x ,
 \label{transported-mode-L2-balance}
\end{equation}
for periodic boundary conditions. Thus, conservation of the
spatially integrated variance requires a volume-preserving flow,
\(\nabla\bcdot U=0\), or an additional approximation that neglects
horizontal compressibility. This condition is not automatic for a
potential flow.

Under a geometrical-optics approximation, the phase satisfies the
same transport equation as \(\varphi_j\). Its local wavevector evolves
along a characteristic according to
\begin{equation}
 \frac{\dif k}{\dif t}
 =
 -\epsilon_\mu(\nabla U)^{\rm tr}k .
 \label{transported-wavevector-ray}
\end{equation}
Writing \(\widehat k=k/|k|\) and denoting by \(S\) the symmetric part
of \(\nabla U\), the radial wavenumber satisfies
\begin{equation}
 \frac{\dif{\mathcal k}}{\dif t}
 =
 -\epsilon_\mu{\mathcal k}\,
 \widehat k\bcdot S\widehat k .
 \label{transported-radial-wavenumber}
\end{equation}
The sign of the radial transfer is therefore determined by the strain
sampled along each ray; it is not fixed by the transport equation
alone.

To obtain an isotropic one-dimensional scaling, we now make two
additional phenomenological assumptions: angular averaging can be
represented by a positive stretching rate
\(\Sigma({\mathcal k})\), and the radial variance flux is constant.
Thus,
\begin{equation}
 \dot{\mathcal k}
 =
 \epsilon_\mu\Sigma({\mathcal k}){\mathcal k},
 \qquad
 \partial_tE_\varphi
 +\partial_{\mathcal k}
 \left(\dot{\mathcal k}E_\varphi\right)
 =0 .
 \label{isotropic-transported-spectrum}
\end{equation}
If \(\Sigma({\mathcal k})\sim{\mathcal k}^s\), a constant variance
flux \(\Pi_\varphi\) gives
\begin{equation}
 E_\varphi({\mathcal k})
 =
 \frac{\Pi_\varphi}
 {\epsilon_\mu\Sigma({\mathcal k}){\mathcal k}}
 \sim
 {\mathcal k}^{-(1+s)} .
 \label{transported-potential-spectrum}
\end{equation}
Equation \eqref{transported-potential-spectrum} is consequently an
isotropic strain closure, rather than a direct consequence of
\eqref{transported-mode-scaling-equation}.

The Airy carrier does not change the equal-time variance of the OU
envelope. If
\begin{equation}
 \sigma_{\mathcal k}\sim{\mathcal k}^{-\alpha},
 \qquad
 \gamma_{\mathcal k}\sim{\mathcal k}^{\beta},
 \qquad
 \Exp|W_{\mathcal k}|^2
 =
 \frac{\sigma_{\mathcal k}^2}{2\gamma_{\mathcal k}},
 \label{OU-transported-power-laws}
\end{equation}
and correlations between the envelope and the transported spatial
mode are neglected at leading order, then
\begin{equation}
 E_{\varphi^\tau}({\mathcal k})
 \sim
 {\mathcal k}^{-(1+s+2\alpha+\beta)} .
 \label{transported-noise-potential-spectrum}
\end{equation}
The corresponding surface horizontal-velocity spectrum is
\begin{equation}
 E_{u_h^\tau}({\mathcal k})
 =
 {\mathcal k}^2E_{\varphi^\tau}({\mathcal k})
 \sim
 {\mathcal k}^{1-s-2\alpha-\beta}.
 \label{transported-surface-velocity-spectrum}
\end{equation}

It is important to distinguish this surface-velocity variance from
Airy wave action and energy. With the normalization used above,
\begin{equation}
 N_{\varphi^\tau}({\mathcal k})
 =
 \frac{\omega_{\mathcal k}}{g}
 E_{\varphi^\tau}({\mathcal k}),
 \qquad
 E_{\rm A}({\mathcal k})
 =
 \omega_{\mathcal k}N_{\varphi^\tau}({\mathcal k})
 =
 G_{\mathcal k}[0]E_{\varphi^\tau}({\mathcal k}).
 \label{potential-action-energy-relations}
\end{equation}
For deep-water gravity waves,
\(\omega_{\mathcal k}\sim{\mathcal k}^{1/2}\) and
\(G_{\mathcal k}[0]\sim{\mathcal k}\), so that
\begin{align}
 N_{\varphi^\tau}({\mathcal k})
 &\sim
 {\mathcal k}^{-1/2-s-2\alpha-\beta},
 \label{transported-action-spectrum}\\
 E_{\rm A}({\mathcal k})
 &\sim
 {\mathcal k}^{-s-2\alpha-\beta}.
 \label{transported-Airy-energy-spectrum}
\end{align}
The carrier phase affects temporal resonance through
\eqref{effective-Airy-correlation-time}, but not these equal-time
relations. A comparison with a Kolmogorov--Zakharov spectrum is
meaningful only after using the same convention for modal versus
shell-integrated spectra and after identifying a conserved flux.

\paragraph{Scaling of the bath-mediated quadratic interaction.}
We finally consider the transported-mode contribution in
\eqref{kinetic-equ-continuous-colored}. For the scaling analysis, we
impose the quartet-wise common-rate condition
\eqref{common-response-rate-condition}. Under this condition, the
general three-kernel operator \eqref{bath-mediated-collision-operator}
reduces to a single-kernel gain--loss operator that is quadratic in the
resolved action, and hence quartic in the resolved wave amplitudes. In
the long-correlation, narrow-band limit
\eqref{narrow-band-resonance-limit}, it becomes
\begin{multline}
 ({\cal Q}_{\cal R}N)_k
 =
 4\pi\epsilon_\mu^2
 \int_{\mathbb R^{3d}}
 {\cal A}_\ell |V_{k\ell pq}|^2
 \left[N_pN_q-N_k(N_p+N_q)\right]
 \\
 {}\times
 \delta^{(d)}(p+q-k-\ell)
 \delta(\omega_p+\omega_q-\omega_k-\omega_\ell)
 \,\dif p\,\dif q\,\dif\ell ,
 \label{resonant-bath-operator-scaling}
\end{multline}
where
\begin{equation}
 {\cal A}_\ell
 :=
 D_\ell|c_\ell|^2
 \label{carrier-spectral-density}
\end{equation}
is the prescribed carrier spectral density. The factor
\({\cal A}_\ell\) already contains the variance of the unresolved
mode. Accordingly, the loss term is
\(-N_k(N_p+N_q)\), not \(-N_kN_\ell\).

The following dimensional estimate applies only to
\eqref{resonant-bath-operator-scaling}; a finite-width OU spectrum
introduces the additional scale \(\gamma_\ell\). In deep water,
\begin{equation}
 V_{\lambda k,\lambda\ell,\lambda p,\lambda q}
 =
 \lambda^3V_{k\ell pq},
 \label{vertex-homogeneity}
\end{equation}
and hence \(|V|^2\) has degree \(6\). In two horizontal dimensions,
the three wavenumber integrations have degree \(6\), the momentum
Dirac distribution has degree \(-2\), and the frequency Dirac
distribution has degree \(-1/2\). Suppose that
\begin{equation}
 {\cal A}_{\lambda\ell}
 =
 \lambda^{-a}{\cal A}_\ell,
 \qquad
 N_{\lambda k}
 =
 \lambda^{-x}N_k .
 \label{quartic-homogeneity-assumptions}
\end{equation}
Then
\begin{equation}
 ({\cal Q}_{\cal R}N)_{\mathcal k}
 \sim
 \epsilon_\mu^2{\mathcal k}^{19/2-a-2x},
 \qquad
 \tau_{\cal R}^{-1}({\mathcal k})
 \sim
 \epsilon_\mu^2{\mathcal k}^{19/2-a-x}.
 \label{bath-operator-homogeneity}
\end{equation}
If \({\cal A}_{\mathcal k}\) is a two-dimensional isotropic density
and
\(E_{\varphi^\tau}({\mathcal k})
=2\pi{\mathcal k}{\cal A}_{\mathcal k}\), then
\eqref{transported-noise-potential-spectrum} gives
\begin{equation}
 a=2+s+2\alpha+\beta .
 \label{carrier-density-exponent}
\end{equation}

Equation \eqref{bath-operator-homogeneity} provides the intrinsic
quartic transfer-rate estimate and remains valid when a stationary
resolved--unresolved equilibrium is imposed. It does not, by itself,
determine \(x\). Moreover, the prescribed-bath operator is not
conservative for the resolved field:
\begin{equation}
 \int_{\mathbb R^2}({\cal Q}_{\cal R}N)_k\,\dif k
 \neq0,
 \qquad
 \int_{\mathbb R^2}\omega_k
 ({\cal Q}_{\cal R}N)_k\,\dif k
 \neq0
 \label{open-quartic-nonconservation}
\end{equation}
in general. The carrier index \(\ell\) is not assigned an adjoint
collision equation, and the permutation symmetries needed to cancel
the resolved action and energy exchanges are therefore absent. The
corrected bracket
\(N_pN_q-N_k(N_p+N_q)\) supplies a gain--loss structure but does not
restore these conservation laws.

The carrier is nevertheless not necessarily frozen for all time.
Its spectrum is prescribed through its initial condition and the OU
parameters, while its spatial correlation modes are transported and
deformed by the resolved flow. The carrier spectrum
\({\cal A}_\ell(t)\) may therefore adapt slowly. We use this property
to introduce an additional equilibrium closure between the quartic
exchange and the spectral flux carried by the unresolved modes.

To state this closure without mixing modal and radial spectra, define
the shell-integrated resolved action and carrier action by
\begin{equation}
 {\mathcal N}_{\rm sh}({\mathcal k})
 =
 2\pi{\mathcal k}N({\mathcal k}),
 \qquad
 {\mathcal C}_{N,\rm sh}({\mathcal k})
 =
 2\pi{\mathcal k}
 \frac{\omega_{\mathcal k}}{g}
 {\cal A}_{\mathcal k}.
 \label{resolved-carrier-shell-actions}
\end{equation}
The corresponding shell-integrated quartic exchange is
\begin{equation}
 {\mathcal T}_{{\cal R},\rm sh}({\mathcal k})
 =
 2\pi{\mathcal k}
 ({\cal Q}_{\cal R}N)_{\mathcal k}.
 \label{shell-quartic-exchange}
\end{equation}
An augmented mean budget can then be written
\begin{align}
 \partial_t{\mathcal N}_{\rm sh}
 +\partial_{\mathcal k}\Pi_N
 &=
 {\mathcal T}_{{\cal R},\rm sh},
 \label{resolved-action-exchange-budget}\\
 \partial_t{\mathcal C}_{N,\rm sh}
 +\partial_{\mathcal k}\Pi_C
 &=
 -{\mathcal T}_{{\cal R},\rm sh}
 +{\mathcal F}_{\rm OU}
 -{\mathcal D}_{\rm OU}.
 \label{carrier-action-exchange-budget}
\end{align}
The second equation constitutes an additional exchange-budget closure
for the transported bath; it does not follow from the prescribed OU
process alone. In a stationary range in which OU injection and damping
balance, this closure gives
\begin{equation}
 {\mathcal T}_{{\cal R},\rm sh}
 =
 -\partial_{\mathcal k}\Pi_C,
 \qquad
 \partial_{\mathcal k}
 \left(\Pi_N+\Pi_C\right)=0.
 \label{resolved-carrier-flux-equilibrium}
\end{equation}
Thus the total flux of the augmented mean budget is constant, even
though \({\cal Q}_{\cal R}\) is not conservative for the resolved
waves. A scale-independent exchange efficiency may be included in
\eqref{resolved-carrier-flux-equilibrium}; it changes the amplitude
but not the exponent derived below. Its magnitude must be chosen so
that the carrier-flux convergence and the quartic term occur on the
same kinetic time scale.

We now estimate the stationary spectrum selected by
\eqref{resolved-carrier-flux-equilibrium}. From
\eqref{bath-operator-homogeneity},
\begin{equation}
 {\mathcal T}_{{\cal R},\rm sh}({\mathcal k})
 \sim
 \epsilon_\mu^2
 {\mathcal k}^{21/2-a-2x}.
 \label{shell-quartic-homogeneity}
\end{equation}
On the other hand,
\({\mathcal C}_{N,\rm sh}\sim{\mathcal k}^{3/2-a}\), and the strain-induced
radial transport velocity satisfies
\(\dot{\mathcal k}\sim{\mathcal k}^{1+s}\). Hence
\begin{equation}
 \Pi_C
 =
 \dot{\mathcal k}\,{\mathcal C}_{N,\rm sh}
 \sim
 {\mathcal k}^{5/2+s-a},
 \qquad
 \partial_{\mathcal k}\Pi_C
 \sim
 {\mathcal k}^{3/2+s-a}.
 \label{carrier-action-flux-scaling}
\end{equation}
Matching the homogeneities in
\eqref{resolved-carrier-flux-equilibrium} gives
\begin{equation}
 \frac{21}{2}-a-2x
 =
 \frac32+s-a,
 \qquad
 x=\frac{9-s}{2}.
 \label{equilibrium-action-exponent}
\end{equation}
The dependence on \(a\), and therefore on the OU exponents
\(\alpha\) and \(\beta\), cancels because the same transported
carrier spectrum determines both the quartic kernel and the supplied
flux. This cancellation is a consequence of the flux-matching
closure, not a universal property of the open collision operator.
For \(s=4/3\), one obtains \(x=23/6\), whereas \(s=1\) gives
\(x=4\). These values coincide respectively with the usual modal
wave-action exponents of the inverse-action and direct-energy gravity
wave cascades, but here they are conditional equilibrium scalings.

The same exponent follows from a consistent energy balance. Indeed,
the carrier shell energy and its flux scale as
\begin{equation}
 {\mathcal C}_{E,\rm sh}
 =
 2\pi{\mathcal k}G_{\mathcal k}[0]
 {\cal A}_{\mathcal k}
 \sim{\mathcal k}^{2-a},
 \qquad
 \Pi_C^E
 =
 \dot{\mathcal k}\,{\mathcal C}_{E,\rm sh}
 \sim{\mathcal k}^{3+s-a}.
 \label{carrier-energy-flux-scaling}
\end{equation}
The resolved shell-energy exchange
\(2\pi{\mathcal k}\omega_{\mathcal k}{\cal Q}_{\cal R}\)
scales as
\({\mathcal k}^{11-a-2x}\); balancing it with
\(\partial_{\mathcal k}\Pi_C^E
\sim{\mathcal k}^{2+s-a}\) again gives
\(x=(9-s)/2\).
The action and energy calculations therefore give the same
homogeneity, but they should be understood as alternative mean-budget
closures. Simultaneous exact conservation of both quantities would
require the reciprocal carrier collision equation discussed below.

The quartic transfer-rate estimate itself remains
\eqref{bath-operator-homogeneity}. After using
\eqref{carrier-density-exponent} and
\eqref{equilibrium-action-exponent}, it becomes
\begin{equation}
 \tau_{\cal R}^{-1}({\mathcal k})
 \sim
 \epsilon_\mu^2
 {\mathcal k}^{3-s/2-2\alpha-\beta}.
 \label{equilibrium-quartic-rate}
\end{equation}
Equivalently, the carrier surface-velocity variance per logarithmic
wavenumber interval satisfies, up to numerical constants,
\begin{equation}
 \sigma_u^2({\mathcal k})
 \sim
 {\mathcal k}^4{\cal A}_{\mathcal k}.
 \label{carrier-velocity-variance-density}
\end{equation}
The intrinsic resonant quartic rate can therefore also be written,
at the level of spectral homogeneity,
\begin{equation}
 \tau_{\cal R}^{-1}({\mathcal k})
 \sim
 \epsilon_\mu^2
 {\mathcal k}^{11/2}
 \sigma_u^2({\mathcal k})N({\mathcal k}).
 \label{quartic-rate-velocity-form}
\end{equation}
Under \eqref{equilibrium-action-exponent}, this becomes
\(\tau_{\cal R}^{-1}
\sim\epsilon_\mu^2
{\mathcal k}^{1+s/2}\sigma_u^2({\mathcal k})\).
Dimensional prefactors involving \(g\) depend on the normalization of
the Airy amplitudes and have been suppressed in this homogeneity
estimate.
Its equality with the diffusive rate in
\eqref{scale-local-variable-D} is not automatic; it requires an
additional relation among the strain, OU-memory, and scattering
scalings.

\paragraph{Production--dissipation decomposition.}
For clarity, the bath-mediated collision operator may be written
\begin{equation}
 ({\cal Q}_{\cal R}N)_k
 =
 P_N(k)-\Gamma_N(k)N_k ,
 \label{correct-production-loss-form}
\end{equation}
where, in the colored-noise formulation,
\begin{align}
 P_N(k)
 &=
 \epsilon_\mu^2
 \int_{\mathbb R^{3d}}
 {\cal K}_{k\ell pq}N_pN_q
 \delta^{(d)}(p+q-k-\ell)
 \,\dif p\,\dif q\,\dif\ell ,
 \label{correct-production-term}\\
 \Gamma_N(k)
 &=
 \epsilon_\mu^2
 \int_{\mathbb R^{3d}}
 {\cal K}_{k\ell pq}(N_p+N_q)
 \delta^{(d)}(p+q-k-\ell)
 \,\dif p\,\dif q\,\dif\ell .
 \label{correct-loss-rate}
\end{align}
Both \(P_N\) and \(\Gamma_N\) are nonnegative. For \(N_k>0\), the
identity
\begin{equation}
 ({\cal Q}_{\cal R}N)_k
 =
 \gamma_{\rm eff}(k)N_k,
 \qquad
 \gamma_{\rm eff}(k)
 =
 \frac{P_N(k)}{N_k}-\Gamma_N(k),
 \label{effective-growth-identity}
\end{equation}
is useful diagnostically, but it does not make the interaction local.
For scale-local quartets, both terms have the same homogeneity,
\begin{equation}
 \frac{P_N({\mathcal k})}{N({\mathcal k})}
 \sim
 \Gamma_N({\mathcal k})
 \sim
 \epsilon_\mu^2
 {\mathcal k}^{19/2-a-x}.
 \label{growth-loss-common-scaling}
\end{equation}
Thus production and loss act on the quartic time scale
\eqref{bath-operator-homogeneity}. Their difference, rather than
either contribution separately, determines wave growth.

\paragraph{Nonlocal wave-growth mechanism.}
For an individual quartet, the sign of its contribution is controlled
by
\begin{equation}
 {\cal R}_{k\ell pq}
 :=
 \frac{N_pN_q}
 {N_k(N_p+N_q)}.
 \label{quartet-growth-ratio}
\end{equation}
It contributes to growth of mode \(k\) when
\({\cal R}_{k\ell pq}>1\), and to loss when
\({\cal R}_{k\ell pq}<1\). The carrier spectrum and vertex multiply
both terms by the same nonnegative rate
\({\cal K}_{k\ell pq}\); they therefore control the strength and
selection of the participating quartets, while the resolved spectra
in \eqref{quartet-growth-ratio} determine the sign of each
contribution.

Assume \(N_r\sim r^{-x}\), with \(x>0\). Several representative scale
configurations illustrate the resulting imbalance:
\begin{itemize}
\item If \(k\ll K\) and
      \(p\sim q\sim\ell\sim K\), then
\[
 {\cal R}_{k\ell pq}
 \sim
 \frac12\left(\frac{k}{K}\right)^x
 \ll1.
\]
Interactions with much shorter resolved and unresolved modes
therefore tend to remove action from the long resolved mode \(k\).

\item If \(k\ll K\), \(p\sim k\), and
      \(q\sim\ell\sim K\), then
\[
 {\cal R}_{k\ell pq}
 \sim
 \left(\frac{k}{K}\right)^x
 \ll1.
\]
With the corrected loss \(N_k(N_p+N_q)\), this mixed configuration is
loss-dominated rather than neutral: the contribution \(N_kN_p\)
dominates the loss.

\item If \(p\sim q\sim\kappa<k\), then
\[
 {\cal R}_{k\ell pq}
 \sim
 \frac12
 \left(\frac{k}{\kappa}\right)^x.
\]
Consequently, two energetic lower-wavenumber modes feed mode \(k\)
when
\[
 \kappa<2^{-1/x}k,
\]
provided that \(\ell\) and the interaction geometry satisfy the
wavenumber and frequency matching conditions. This is the
nonlocal wave-growth configuration.
\end{itemize}
These estimates are conditional on the resonance manifold
\[
 p+q=k+\ell,
 \qquad
 \omega_p+\omega_q=\omega_k+\omega_\ell .
\]
In particular, the asymptotic configuration
\(p,q\ll k\) with \(\ell\sim k\) is excluded by the positive-frequency
deep-water resonance, because its left-hand frequency is too small.
Growth can nevertheless occur for moderately lower \(p\) and \(q\)
on the exact resonance manifold. For a finite-correlation OU carrier,
off-resonant configurations also contribute, but are reduced by the
Lorentzian factor
\(\widehat R_\ell(\Omega)
\propto(\gamma_\ell^2+\Omega^2)^{-1}\).

The sign of the integrated growth rate
\(\gamma_{\rm eff}(k)\) consequently cannot be inferred from scale
ordering alone: it is the weighted balance of all admissible quartets.
Nevertheless, \eqref{quartet-growth-ratio} exhibits a clear
turbulence-mediated growth mechanism. Energetic lower-wavenumber
resolved modes provide the quadratic production \(N_pN_q\), while
the transported unresolved spectrum
\({\cal A}_\ell\) and the OU temporal spectrum select and amplify the
coupling. The quartic term therefore cannot create wave action from
an entirely vanishing resolved spectrum. It can, however, populate an
initially empty mode \(k\) from nonzero modes \(p\) and \(q\), and can
amplify a wave field seeded by initial waves or by the additive
stochastic forcing.

Writing
\(\partial_tN_k=\gamma_{\rm eff}(k)N_k\) is analogous to a
Miles-type growth law, but the physical mechanism is different.
Miles growth is a local linear instability associated with a
critical level in the wind profile. Here
\(\gamma_{\rm eff}\) is a nonlinear, spectrum-dependent and
intrinsically nonlocal rate generated by stochastic transport and
resolved--unresolved wave coupling. It consequently provides an
alternative mechanism by which turbulence and unresolved waves can
produce net wave growth.

The strain exponent controls the strength of this mechanism through
the transported carrier density
\[
 {\cal A}_{\mathcal k}
 \sim
 {\mathcal k}^{-a},
 \qquad
 a=2+s+2\alpha+\beta.
\]
Increasing \(s\) steepens the carrier spectrum and weakens the
contribution of high-wavenumber unresolved modes, whereas smaller
\(s\) enhances their coupling. The actual growth or decay remains
controlled by the resonance geometry and the production--loss
imbalance in \eqref{quartet-growth-ratio}.

The equilibrium closure
\eqref{resolved-action-exchange-budget}--\eqref{carrier-action-exchange-budget}
does not require the carrier to become a fully Hamiltonian wave
field. It is compatible with a bath initialized through
\({\cal A}_\ell(0)\), maintained temporally by the OU process, and
adapted spatially through transport of its correlation modes. It is,
however, a closure of the mean budgets and not a proof that the
quartic operator is conservative. Exact conservation for an enlarged
resolved--carrier system would require an explicit reciprocal
collision equation for the carrier. In a closed Hamiltonian model,
this leads instead to the mixed cubic bracket
\eqref{mixed-Hasselmann-bracket}. The OU forcing and damping would
still appear as external injection and dissipation in such an
enlarged system.

\subsection{Spectral-peak evolution}

While the preceding analysis characterizes stationary power-law
solutions, many ocean-wave spectra remain narrow-banded. It is
therefore useful to derive an equation for the spectral peak, which
provides a reduced description of the competition between stochastic
scattering, bath-mediated nonlinear interactions, and external
forcing and dissipation. We consider an isotropic modal action density
\(N({\mathcal k},t)\) with a nondegenerate maximum at
\({\mathcal k}_p(t)\):
\begin{equation}
 \partial_{\mathcal k}N({\mathcal k}_p,t)=0,
 \qquad
 \partial_{\mathcal k\mathcal k}N({\mathcal k}_p,t)<0.
 \label{spectral-peak-conditions}
\end{equation}
Differentiating the first relation with respect to time gives the
exact kinematic identity
\begin{equation}
 \dot{\mathcal k}_p
 =
 -\left.
 \frac{\partial_{\mathcal k}(\partial_tN)}
 {\partial_{\mathcal k\mathcal k}N}
 \right|_{{\mathcal k}={\mathcal k}_p}.
 \label{exact-peak-identity}
\end{equation}
For the radial kinetic equation
\begin{equation}
 \partial_tN
 =
 \frac{1}{\mathcal k}
 \partial_{\mathcal k}
 \left({\mathcal k}D({\mathcal k})
 \partial_{\mathcal k}N\right)
 +{\cal Q}_{\cal R}[N]
 +{\cal I}({\mathcal k})
 -2\mu_{\mathcal k}N ,
 \label{radial-full-kinetic-peak}
\end{equation}
where \({\cal I}\) contains external injection not already represented
by the carrier exchange.

\paragraph{Transport-noise scattering.}
Denote the radial diffusion operator by
\[
 {\cal L}_D N
 =
 \frac{1}{\mathcal k}
 \partial_{\mathcal k}
 \left({\mathcal k}D\partial_{\mathcal k}N\right).
\]
It can be expanded as
\begin{equation}
 {\cal L}_D N
 =
 D N''+
 \left(D'+\frac{D}{\mathcal k}\right)N'.
 \label{expanded-radial-diffusion}
\end{equation}
Consequently, its derivative at the peak is
\begin{equation}
 \left.
 \partial_{\mathcal k}
 \left[
 \frac{1}{\mathcal k}\partial_{\mathcal k}
 \left({\mathcal k}D\partial_{\mathcal k}N\right)
 \right]\right|_{{\mathcal k}_p}
 =
 \left(2D'+\frac{D}{\mathcal k}\right)
 N''
 +DN'''
 \bigg|_{{\mathcal k}_p}.
 \label{correct-radial-diffusion-derivative}
\end{equation}
and the corresponding peak velocity is
\begin{equation}
 \dot{\mathcal k}_p^{({\rm sc})}
 =
 -\left(2D'({\mathcal k}_p)
 +\frac{D({\mathcal k}_p)}{{\mathcal k}_p}\right)
 -D({\mathcal k}_p)
 \frac{N'''({\mathcal k}_p)}
 {N''({\mathcal k}_p)} .
 \label{general-diffusive-peak-drift}
\end{equation}
Here and below, primes denote radial derivatives evaluated at the
spectral peak when no argument is displayed.

Under the stronger \(k\)-independent condition
\({\mathcal G}^{\rm A}(k,q)={\mathcal G}_0(q)\), one has \(D=D_0\),
and hence
\begin{equation}
 \dot{\mathcal k}_p^{({\rm sc})}
 =
 -\frac{D_0}{{\mathcal k}_p}
 -D_0\frac{N'''({\mathcal k}_p)}
 {N''({\mathcal k}_p)} .
 \label{constant-D-peak-drift}
\end{equation}
For a locally symmetric peak, the second term vanishes. The remaining
inverse displacement is a consequence of the radial geometry of
two-dimensional diffusion. This radial-geometrical contribution would
not appear in a one-dimensional Cartesian diffusion equation.

For a scale-local diffusivity
\begin{equation}
 D({\mathcal k})
 \simeq
 D_*{\mathcal k}^{m_D},
 \qquad D_*>0,
 \label{local-power-law-diffusivity-peak}
\end{equation}
equation \eqref{general-diffusive-peak-drift} becomes
\begin{equation}
 \dot{\mathcal k}_p^{({\rm sc})}
 =
 -(2m_D+1)\frac{D({\mathcal k}_p)}{{\mathcal k}_p}
 -D({\mathcal k}_p)
 \frac{N'''({\mathcal k}_p)}{N''({\mathcal k}_p)}.
 \label{power-law-D-peak-drift}
\end{equation}
In particular, if the velocity variance and correlation time vary
negligibly across the narrow peak, the estimate
\eqref{scale-local-variable-D} gives
\(D({\mathcal k})\simeq
\sigma_u^2\tau_{\rm eff}{\mathcal k}^4\), and hence
\begin{equation}
 \dot{\mathcal k}_p^{({\rm sc})}
 \simeq
 -9\sigma_u^2\tau_{\rm eff}{\mathcal k}_p^3
 -\sigma_u^2\tau_{\rm eff}{\mathcal k}_p^4
 \frac{N'''({\mathcal k}_p)}{N''({\mathcal k}_p)}.
 \label{k4-diffusive-peak-drift}
\end{equation}
The factor \(9\), rather than \(4\), results from the two derivatives
of the variable diffusivity together with the radial-geometrical term
\(D/{\mathcal k}\). For a locally symmetric peak, diffusion therefore
produces systematic inverse migration both for constant \(D\) and for
the scale-local \({\mathcal k}^4\) law. More generally, this conclusion
holds for a symmetric peak whenever \(m_D>-1/2\).

\paragraph{Bath-mediated production--dissipation contribution.}
From \eqref{correct-production-loss-form}, define the net local
exchange
\begin{equation}
 \Delta_N({\mathcal k})
 :=
 P_N({\mathcal k})
 -\Gamma_N({\mathcal k})N({\mathcal k})
 =
 ({\cal Q}_{\cal R}N)({\mathcal k}).
 \label{peak-production-loss-imbalance}
\end{equation}
Its exact contribution to the peak velocity is
\begin{equation}
 \dot{\mathcal k}_p^{({\cal R})}
 =
 -\left.
 \frac{\partial_{\mathcal k}{\cal Q}_{\cal R}[N]}
 {N''}
 \right|_{{\mathcal k}_p}.
 \label{quartic-peak-contribution}
\end{equation}
Because \({\cal Q}_{\cal R}\) is a nonlocal collision integral, the
sign of \(\Delta_N({\mathcal k}_p)\) determines local wave growth or
loss but does not, by itself, determine the direction of peak motion.
The latter is controlled by
\(\partial_{\mathcal k}{\cal Q}_{\cal R}[N]\).

The earlier production--dissipation interpretation is recovered under
an additional scale-local homogeneity approximation. Factoring the
homogeneous kernel from the collision operator, write
\begin{equation}
 ({\cal Q}_{\cal R}N)({\mathcal k})
 \simeq
 \epsilon_\mu^2{\mathcal k}^{\nu}
 \widehat\Delta_N({\mathcal k}),
 \qquad
 \nu=\frac{19}{2}-a,
 \label{scale-local-peak-collision}
\end{equation}
where \(\widehat\Delta_N\) is the reduced production--loss imbalance
after removal of the common homogeneous prefactor. If the modal
unresolved Airy action satisfies
\(N^{\rm u}_{\mathcal k}\sim{\mathcal k}^{-y}\), then
\({\cal A}_{\mathcal k}=(g/\omega_{\mathcal k})
N^{\rm u}_{\mathcal k}\) gives, in deep water,
\(a=y+1/2\) and therefore \(\nu=9-y\). The simplified assumption
\(y=1+s\) used in the initial scale-local estimate gives
\(\nu=8-s\).

Differentiating \eqref{scale-local-peak-collision} yields
\begin{equation}
 \dot{\mathcal k}_p^{({\cal R})}
 \simeq
 -\epsilon_\mu^2{\mathcal k}_p^{\nu-1}
 \frac{
 \nu\widehat\Delta_N({\mathcal k}_p)
 +{\mathcal k}_p\widehat\Delta_N'({\mathcal k}_p)
 }{N''({\mathcal k}_p)}.
 \label{scale-local-quartic-peak-drift}
\end{equation}
Since \(N''({\mathcal k}_p)<0\), a positive numerator produces
forward peak migration and a negative numerator produces inverse
migration. If
\begin{equation}
 \left|{\mathcal k}_p\widehat\Delta_N'({\mathcal k}_p)\right|
 \ll
 \nu\left|\widehat\Delta_N({\mathcal k}_p)\right|,
 \qquad \nu>0,
 \label{slow-imbalance-peak-condition}
\end{equation}
then production in excess of loss gives forward migration, whereas
loss in excess of production gives inverse migration. For
\(y=1+s\), equation \eqref{scale-local-quartic-peak-drift} reads
\begin{equation}
 \dot{\mathcal k}_p^{({\cal R})}
 \simeq
 -\epsilon_\mu^2{\mathcal k}_p^{7-s}
 \frac{
 (8-s)\widehat\Delta_N({\mathcal k}_p)
 +{\mathcal k}_p\widehat\Delta_N'({\mathcal k}_p)
 }{N''({\mathcal k}_p)}.
 \label{strain-quartic-peak-drift}
\end{equation}
This is the qualified version of the production--dissipation cascade
interpretation: the reduction is not valid when nonlocal resonance
geometry makes \(\widehat\Delta_N\) vary rapidly with
\({\mathcal k}\).

\paragraph{Carrier-flux closure.}
Under the equilibrium closure
\eqref{resolved-carrier-flux-equilibrium}, the transported-carrier
flux closes the bath-mediated interaction rather than supplying a
second independent source. In terms of the shell flux,
\begin{equation}
 {\cal Q}_{\cal R}[N]({\mathcal k})
 =
 -\frac{1}{2\pi{\mathcal k}}
 \partial_{\mathcal k}\Pi_C({\mathcal k}).
 \label{quartic-carrier-flux-closure}
\end{equation}
and hence
\begin{equation}
 \dot{\mathcal k}_p^{({\cal R})}
 =
 \frac{1}{2\pi N''({\mathcal k}_p)}
 \left[
 \frac{\Pi_C''({\mathcal k}_p)}{{\mathcal k}_p}
 -\frac{\Pi_C'({\mathcal k}_p)}{{\mathcal k}_p^2}
 \right].
 \label{carrier-flux-peak-drift}
\end{equation}
Thus the sign of the carrier contribution is determined by the
convergence and radial variation of its shell flux. Equation
\eqref{carrier-flux-peak-drift} replaces
\eqref{quartic-peak-contribution} when the equilibrium closure is
used; adding the two expressions would count the same exchange twice.
Outside this closure, the full nonlocal collision integral must be
retained. A genuinely independent unresolved-wave injection may still
be included in \({\cal I}\).

\paragraph{External forcing and dissipation.}
The remaining contribution is
\begin{equation}
 \dot{\mathcal k}_p^{({\rm ext})}
 =
 -\left.
 \frac{{\cal I}'-2\mu_{\mathcal k}'N}
 {N''}
 \right|_{{\mathcal k}_p}.
 \label{external-peak-contribution}
\end{equation}
The term \(-2\mu_{\mathcal k}N'\) vanishes at the peak. Injection or
dissipation changes the peak position only through its radial
variation; a locally constant damping rate and a forcing profile
symmetric about \({\mathcal k}_p\) do not contribute instantaneously
to the peak velocity.

\paragraph{Resulting peak dynamics and regimes.}
Combining the exact contributions gives
\begin{multline}
 \dot{\mathcal k}_p
 =
 -\left(2D'({\mathcal k}_p)
 +\frac{D({\mathcal k}_p)}{{\mathcal k}_p}\right)
 -D({\mathcal k}_p)
 \frac{N'''({\mathcal k}_p)}{N''({\mathcal k}_p)}
 \\
 {}-
 \frac{
 \partial_{\mathcal k}{\cal Q}_{\cal R}[N]({\mathcal k}_p)
 +{\cal I}'({\mathcal k}_p)
 -2\mu_{{\mathcal k}_p}'N({\mathcal k}_p)
 }{N''({\mathcal k}_p)}.
 \label{complete-spectral-peak-dynamics}
\end{multline}
When the external contribution is negligible over an inertial range,
the relative importance of scattering and bath-mediated interaction
can be measured by
\begin{equation}
 {\mathfrak R}_p
 :=
 \frac{
 \left|\partial_{\mathcal k}{\cal Q}_{\cal R}[N]
 ({\mathcal k}_p)\right|
 }{
 \left|
 \left(2D'+D/{\mathcal k}\right)N''+DN'''
 \right|_{{\mathcal k}_p}
 }.
 \label{peak-regime-ratio}
\end{equation}
This identifies three dynamical regimes:
\begin{itemize}
 \item If \({\mathfrak R}_p\ll1\), scattering diffusion dominates.
 For a locally symmetric peak and either constant \(D\) or the
 scale-local \({\mathcal k}^4\) diffusivity, the peak migrates toward
 smaller wavenumbers.

 \item If \({\mathfrak R}_p\gg1\), the bath-mediated interaction
 dominates. Its direction is determined exactly by
 \(\partial_{\mathcal k}{\cal Q}_{\cal R}\). Under
 \eqref{slow-imbalance-peak-condition}, excess production produces
 forward migration, whereas excess loss produces inverse migration.
 This may be interpreted as a cascade-dominated peak evolution when
 the same spectral range also carries a nonzero approximately constant
 action or energy flux.

 \item If \({\mathfrak R}_p={\cal O}(1)\), diffusion and nonlinear
 exchange compete. Their balance may slow, arrest, or reverse the
 spectral-peak migration.
\end{itemize}
Peak migration and cascade should nevertheless be distinguished: a
displacement of the maximum diagnoses spectral evolution, whereas a
cascade additionally requires a nonzero flux across a wavenumber
range. The peak equation therefore supplies a reduced diagnostic of
the competition between the two mechanisms, rather than by itself a
proof of a constant-flux cascade.

Under the leading-order rate balance
\eqref{asymptotically-balanced-Airy-rates}, the robust conclusions are
the variable-coefficient diffusion tensor
\eqref{variable-scattering-diffusion}, the radial flux relation
\eqref{radial-action-flux-variable-D}, and the general peak formula
\eqref{general-diffusive-peak-drift}. A power-law spectrum additionally
requires the scale-local estimate and explicit velocity and memory
scalings in \eqref{scale-local-variable-D}--\eqref{velocity-memory-power-laws}.
For the diffusive branch, the particular exponent \(23/6\) is
obtained only under condition \eqref{condition-for-23-over-6}. For the
interaction branch supplemented by the resolved--carrier equilibrium
closure, the conditional exponent is instead
\(x=(9-s)/2\), and \(x=23/6\) requires \(s=4/3\). Neither result is a
consequence of the leading-order rate balance alone. These stationary
exponents complement, but do not replace, the dynamical regime
classification in \eqref{peak-regime-ratio}.

\section{Numerical comparison between stochastic transport and Hasselmann interactions}

The objective of this section is to compare the characteristic transfer rates
associated with stochastic transport and with the classical Hasselmann
four-wave interaction mechanism under representative deep-water ocean
conditions. The comparison is performed using the asymptotic transfer rates
derived in the previous sections.

For weakly nonlinear gravity waves, the resonant transfer rate associated with
the Hasselmann--Zakharov equation scales as
\[
\tau_{HZ}^{-1}\sim \epsilon_\mu^4\omega_k,
\]
where $\epsilon_\mu$ denotes the wave steepness and
$\omega_k=\sqrt{gk}$.

The stochastic transport mechanism is characterized by the diffusive transfer
rate
\[
\tau_{tr}^{-1}\sim k^2\sigma_u^2\tau_c,
\]
where $\sigma_u$ denotes the rms unresolved velocity and $\tau_c$ its
decorrelation time.

Throughout this section we compare both mechanisms through the dimensionless
ratio

\[
\Gamma
=
\frac{\tau_{HZ}}{\tau_{tr}}
=
\frac{k^2\sigma_u^2\tau_c}
{\epsilon_\mu^4\omega_k}.
\]

Large values ($\Gamma>1$) indicate faster stochastic transport,
whereas $\Gamma<1$ corresponds to a regime dominated by resonant nonlinear
interactions.
The comparison should be interpreted as an ordering of the characteristic
transfer time scales rather than as a term-by-term comparison of the two
kinetic equations.


\subsection{The Regime Diagram: Transport-Dominated Regimes Under Typical Ocean Parameters}
\begin{figure}[htbp]
    \centering
    \includegraphics[width=0.45\textwidth]{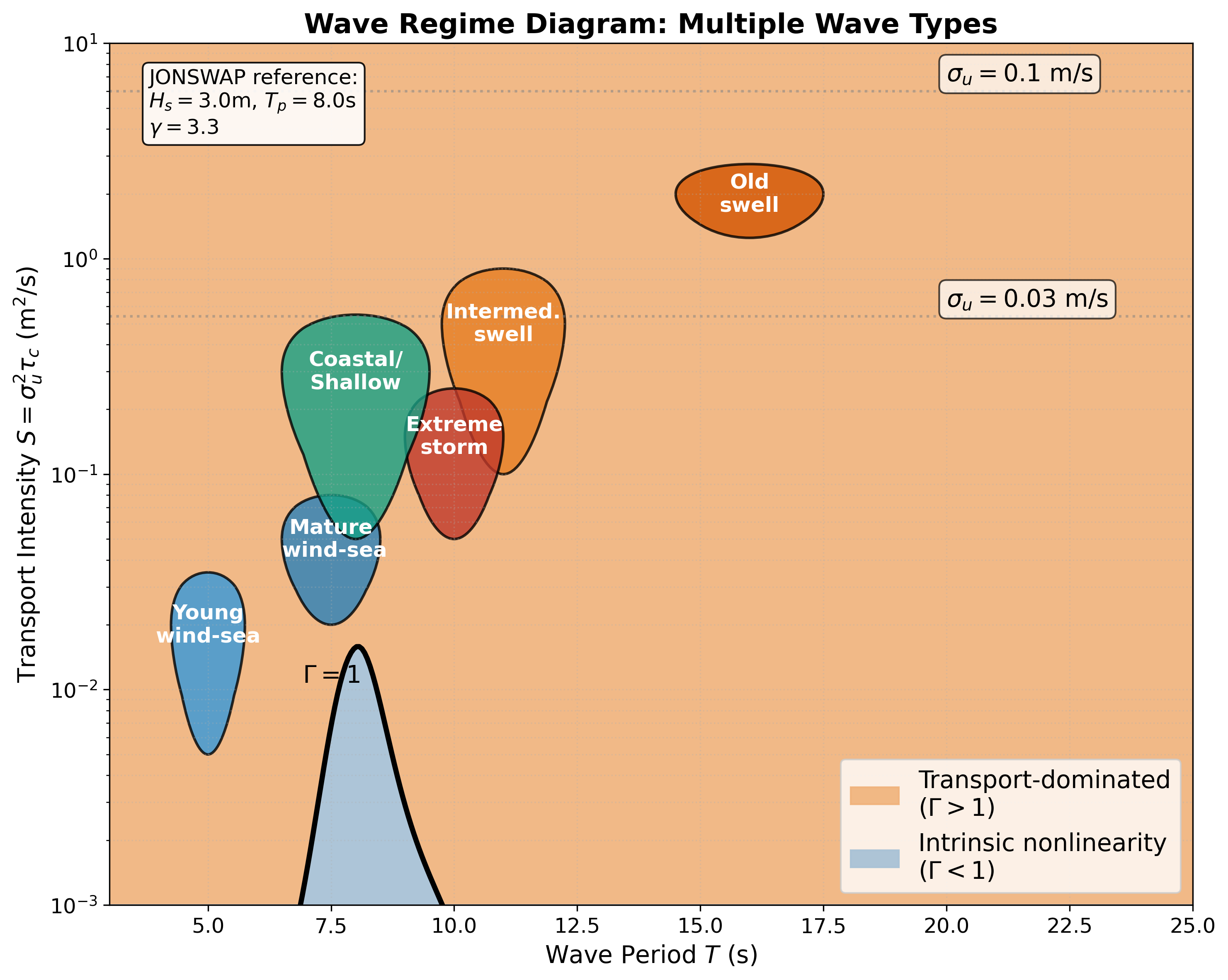} \includegraphics[width=0.45\textwidth]{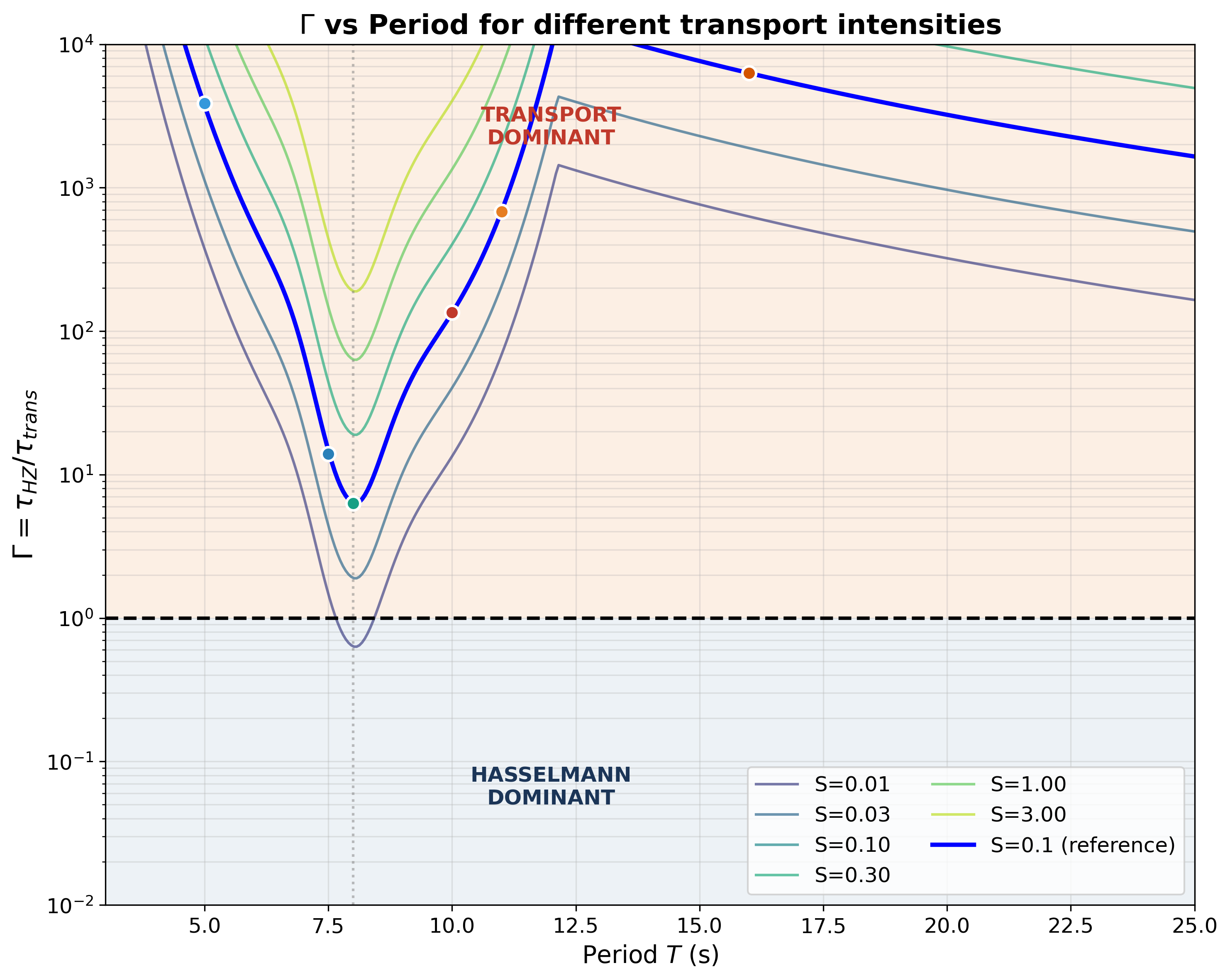}\\
        \includegraphics[width=0.45\textwidth]{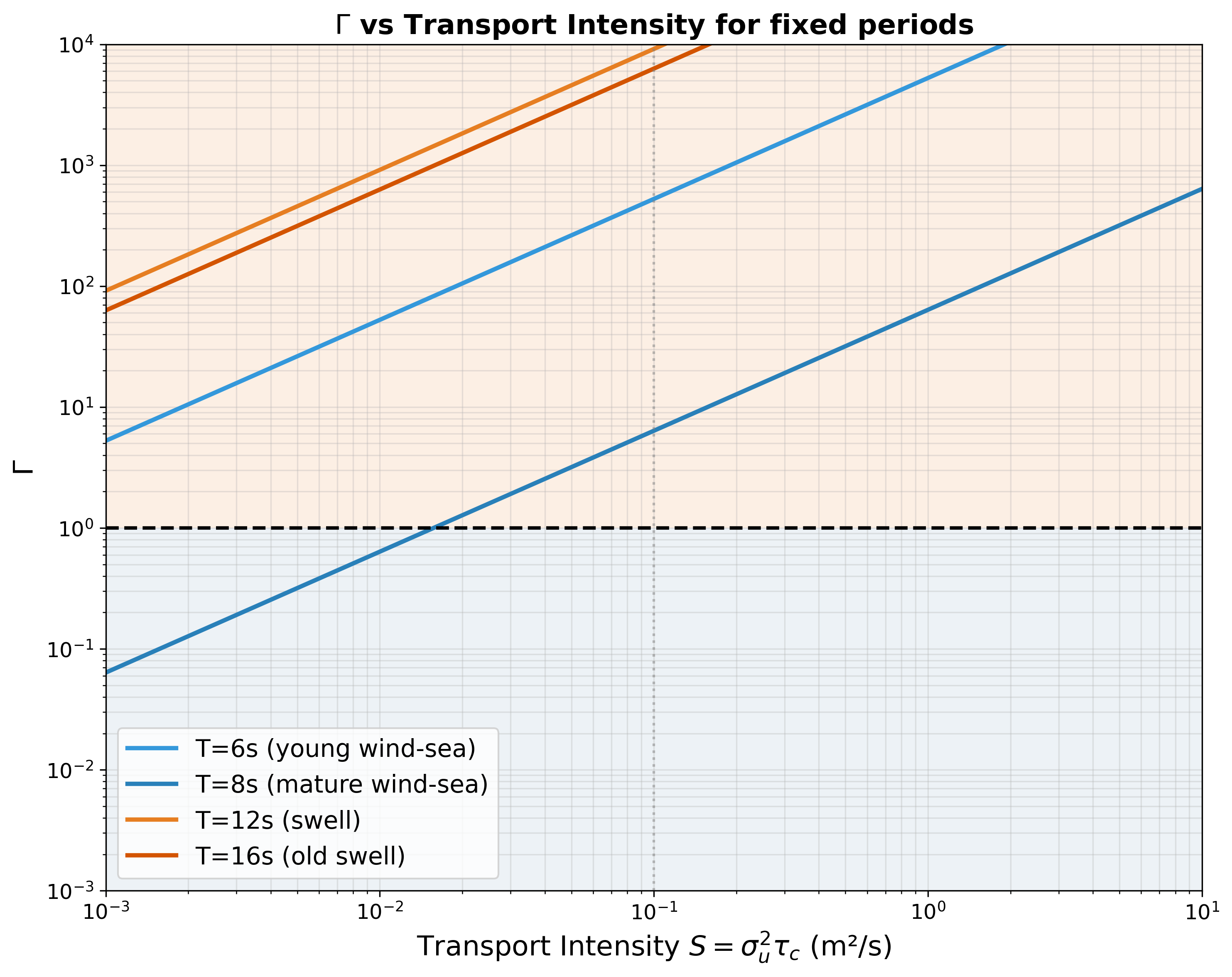}
            \includegraphics[width=0.45\textwidth]{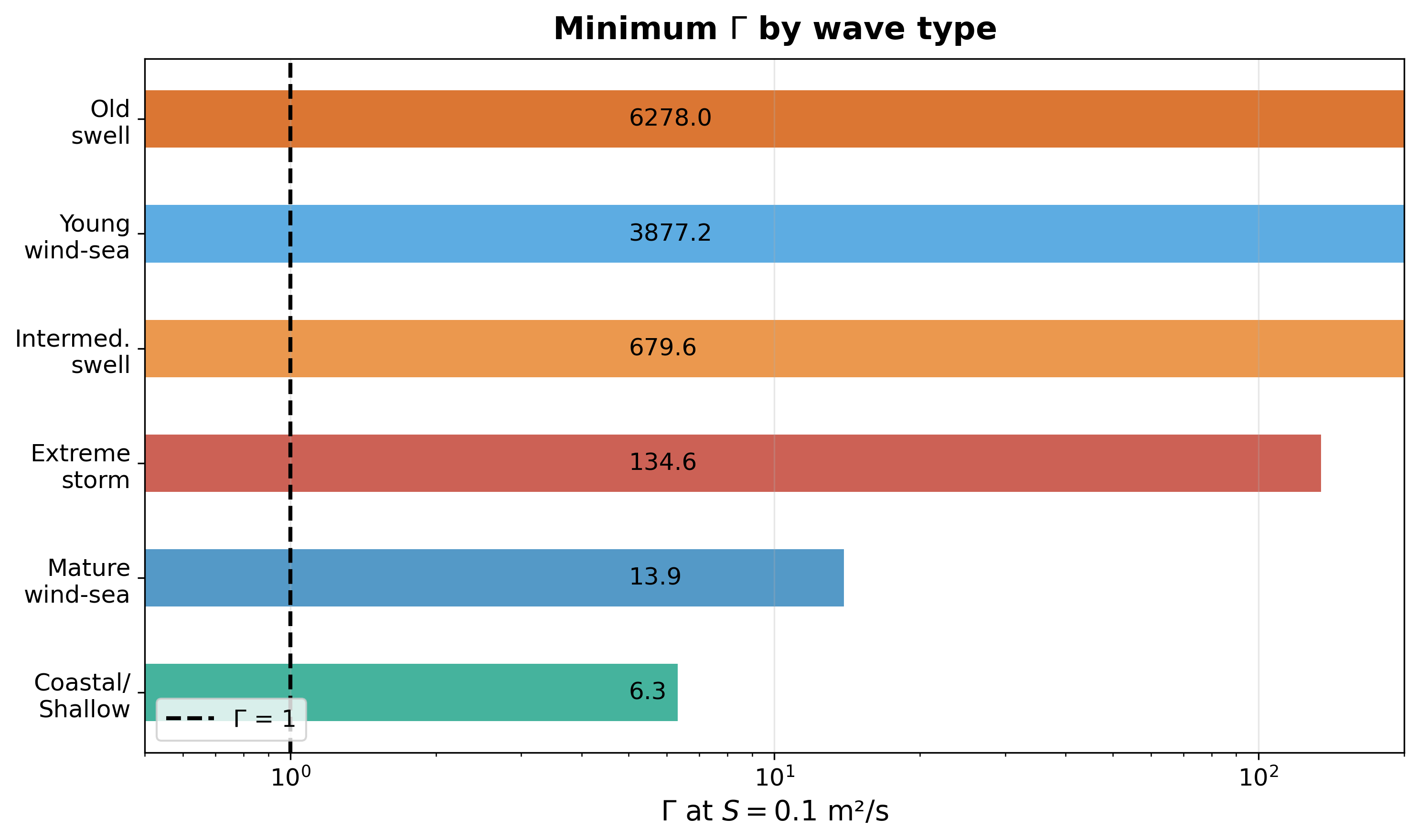}
    \caption{
    \textbf{Regime diagram and quantitative analysis of wave dynamics.}
    (a) Regime diagram in the parameter space of wave period $T$ and transport intensity $S = \sigma_u^2 \tau_c$. The black curve ($\Gamma = 1$) separates the transport-dominated regime ($\Gamma > 1$, orange) from the Hasselmann-dominated regime ($\Gamma < 1$, blue). The horizontal dotted lines indicate constant $\sigma_u$ values for $\tau_c = 600$ s. Colored ellipses show the typical locations of different wave types.  
    (b) $\Gamma$ as a function of period $T$ for multiple fixed $S$ values. The thick blue line corresponds to the reference value $S = 0.1$ m$^2$/s (typical ocean turbulence).  
    (c) $\Gamma$ as a function of $S$ for fixed periods $T = 6, 8, 12, 16$ s, showing the linear scaling $\Gamma \propto S$.  
    (d) Minimum $\Gamma$ values at $S = 0.1$ m$^2$/s for each wave type, all exceeding unity by at least an order of magnitude.
    }
    \label{fig:regime}
\end{figure}
Figure~\ref{fig:regime} summarizes the competition between the two transfer
mechanisms over the parameter space spanned by the wave period
$T$ and the transport intensity $S=\sigma_u^2\tau_c$. The black curve $\Gamma=1$ separates two asymptotic regimes:

\begin{itemize}

\item
$\Gamma<1$: resonant Hasselmann interactions dominate;

\item
$\Gamma>1$: stochastic transport dominates.

\end{itemize}

For representative oceanic values
$S\simeq0.1\,{\rm m^2\,s^{-1}}$
(corresponding for example to
$\sigma_u\simeq0.1\,{\rm m\,s^{-1}}$
and
$\tau_c\simeq10\,{\rm s}$),
all wave periods between approximately
$5$ and $20$ seconds lie in the transport-dominated regime.

Only for substantially weaker unresolved motions
($S\lesssim10^{-2}\,{\rm m^2\,s^{-1}}$)
does the Hasselmann mechanism become competitive.

Figure~\ref{fig:regime}b illustrates the dependence of $\Gamma$
upon the wave period.
A pronounced minimum appears near the spectral peak
($T\simeq8$ s), where the wave steepness is maximal.
Since the resonant interaction rate scales as
$\epsilon_\mu^4$,
even moderate variations of steepness strongly reduce $\Gamma$,
making this region the most favourable for resonant interactions.
Nevertheless, for the representative transport intensity considered here,
the minimum remains above unity.
\subsection{Sensitivity to transport intensity}
The dependence upon unresolved motions is shown in
Figures~\ref{fig:regime}c--d. 

Figure~\ref{fig:regime}c confirms the expected linear dependence $\Gamma\propto \sigma_u^2\tau_c.$
Consequently,
reducing the unresolved velocity variance by one order of magnitude
reduces $\Gamma$ by the same amount.
The transition between both regimes therefore occurs primarily through the
intensity of unresolved transport.

Figure~\ref{fig:regime}d summarizes the minimum values of $\Gamma$
for representative wave types.
Young wind-seas exhibit the smallest ratios because of their relatively
large steepness,
whereas old swell reaches values exceeding
$10^3$ owing to the weakness of nonlinear resonant interactions.
For all wave classes considered,
the minimum remains larger than unity for the reference transport intensity.

\subsection{Spectral dependence}
\begin{figure}[htbp]
\centering
    \includegraphics[width=0.45\textwidth]{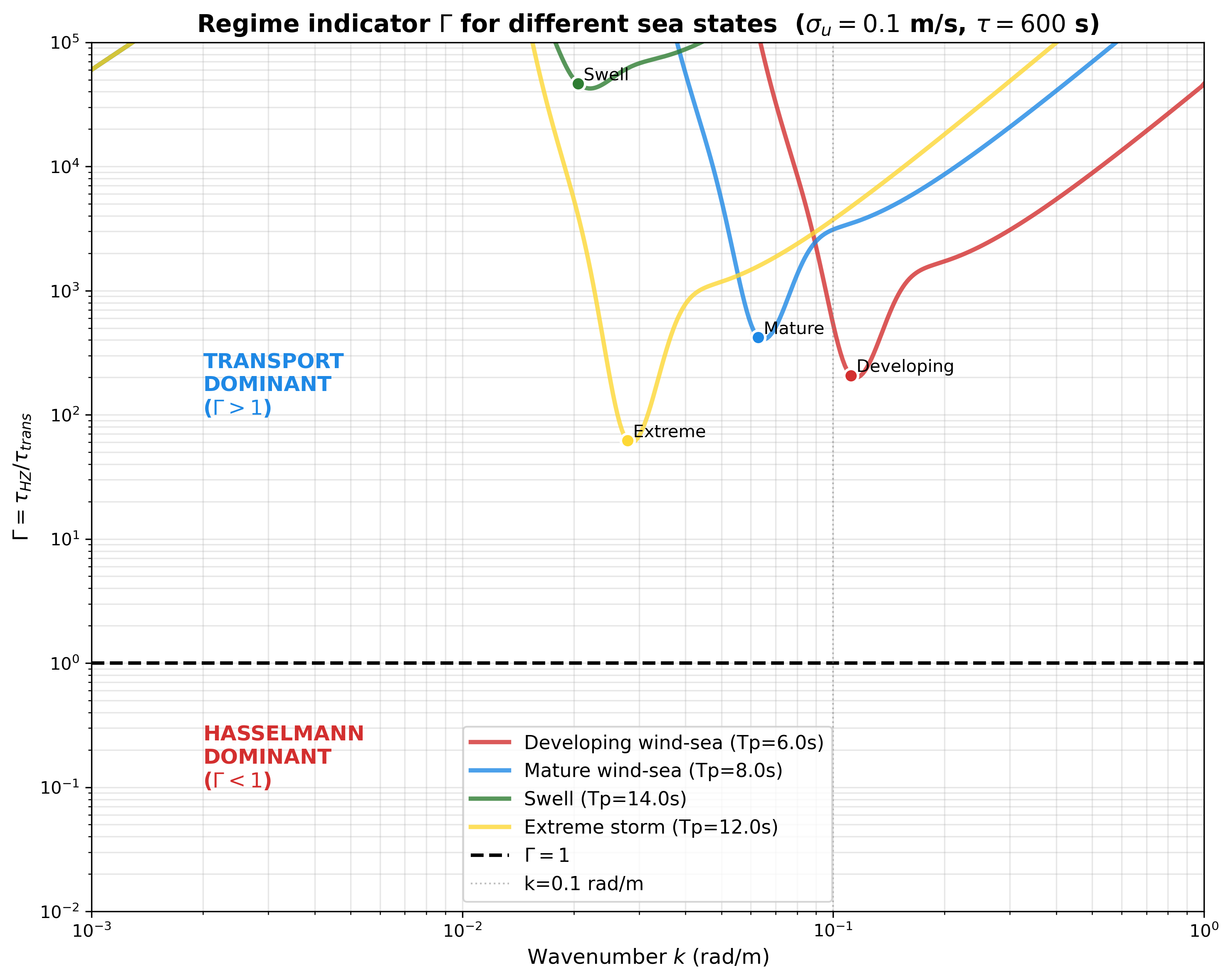} \includegraphics[width=0.45\textwidth]{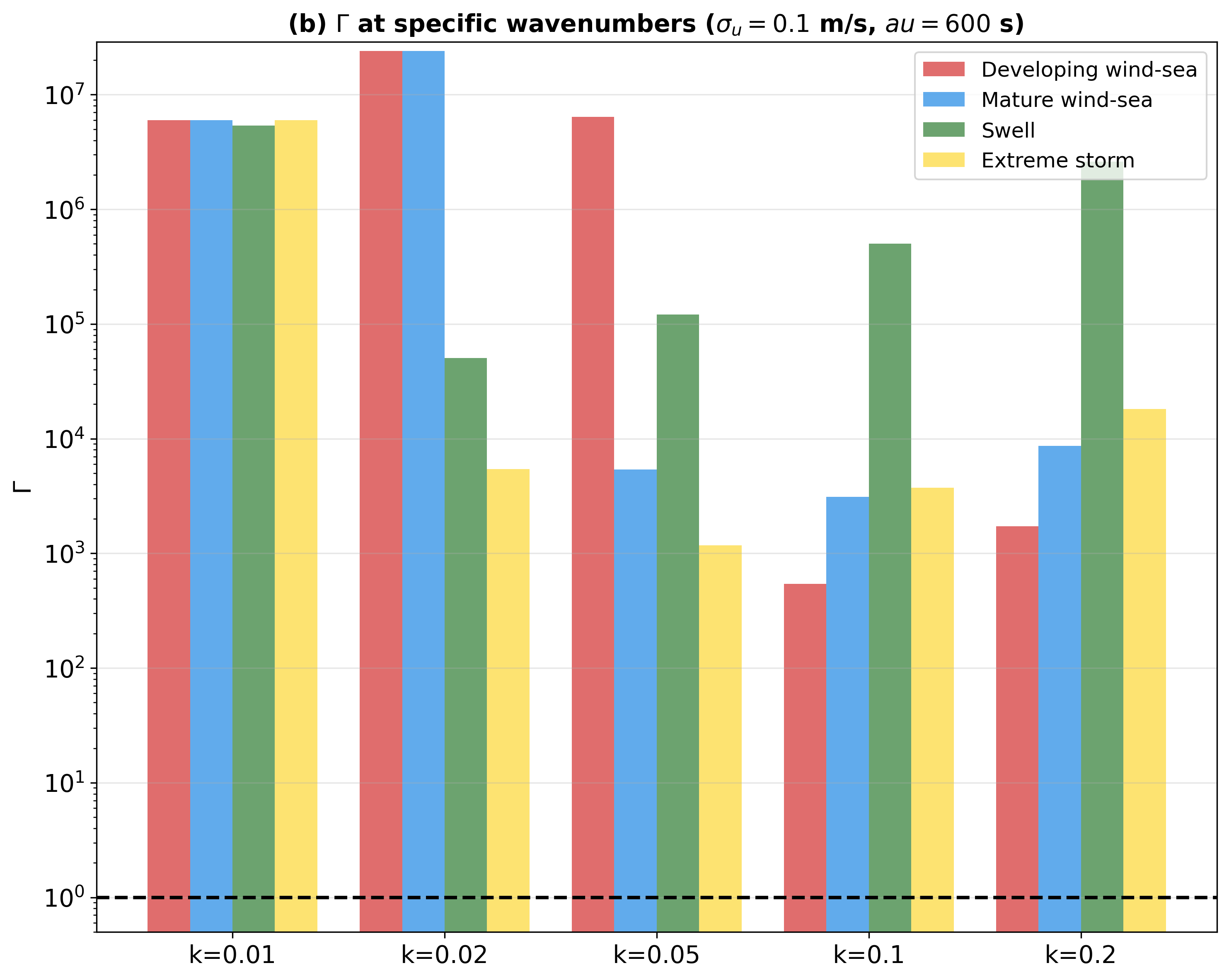}\\
        \includegraphics[width=0.45\textwidth]{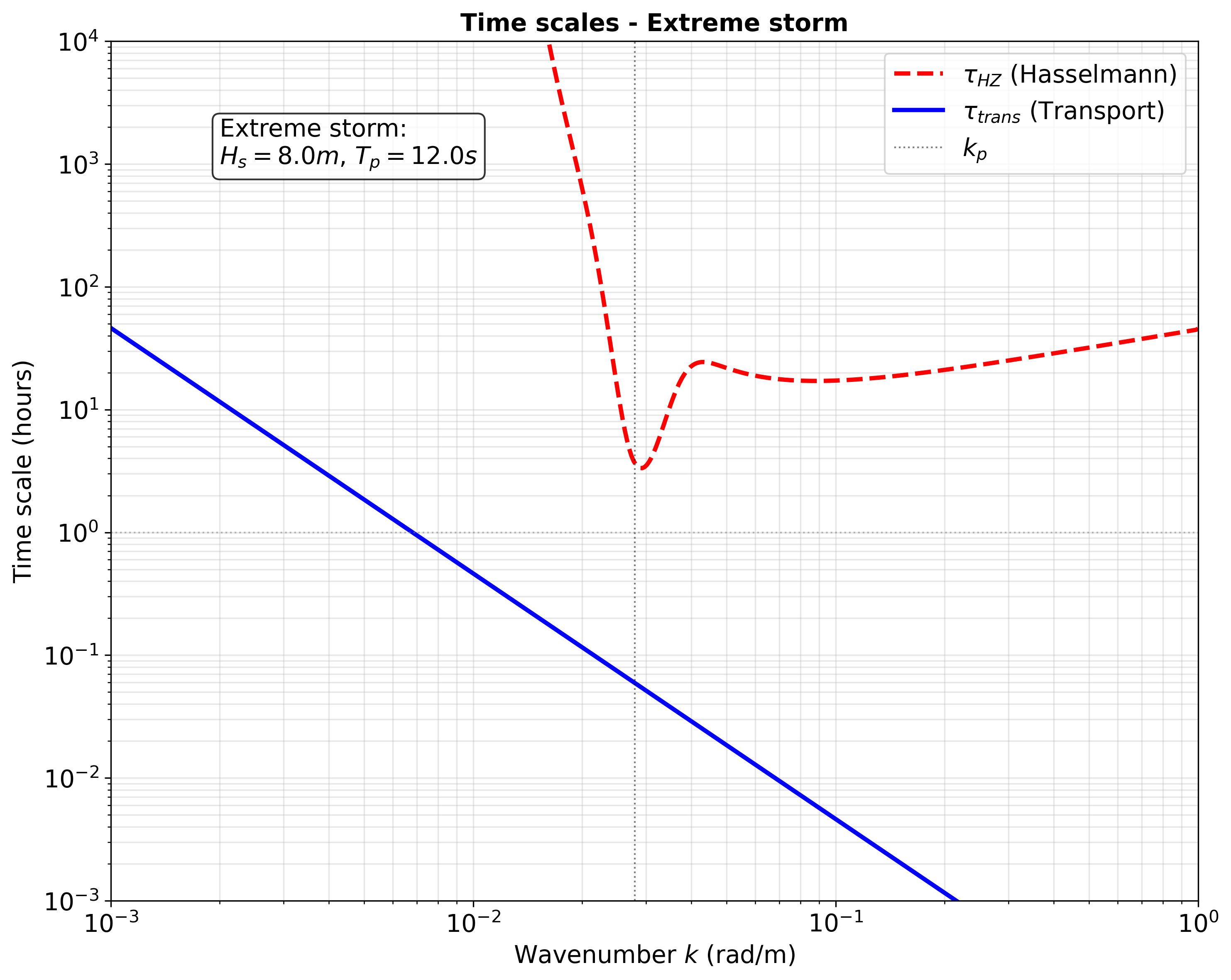}\\~\\
            \includegraphics[width=\textwidth]{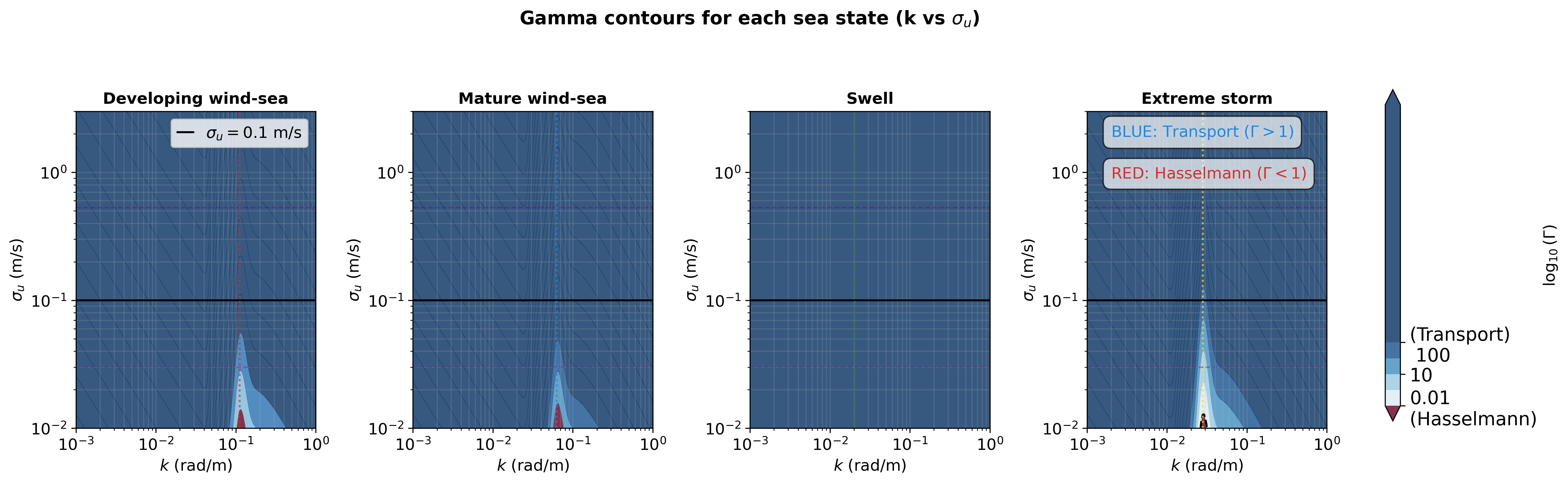}
\caption{
\textbf{Multi-panel regime analysis of ocean wave dynamics.}
(a) Regime indicator $\Gamma = \tau_{HZ}/\tau_{trans}$ as a function of wavenumber $k$ for different sea states. The horizontal dashed line at $\Gamma = 1$ separates the transport-dominated regime ($\Gamma > 1$, blue) from the Hasselmann-dominated regime ($\Gamma < 1$, red). For the sea states considered, the curves remain in the $\Gamma>1$ region across the resolved wavenumber range.
(b) Bar chart showing $\Gamma$ values at specific wavenumbers for each sea state. Even at the highest wavenumber ($k = 0.2$ rad/m), $\Gamma$ remains well above unity.  
(c) Time scale comparison for an extreme storm ($H_s = 8$ m, $T_p = 12$ s). The Hasselmann time scale $\tau_{HZ}$ (red dashed) remains larger than the transport time scale $\tau_{trans}$ (blue solid) over the wavenumber range considered.  
(d) Gamma contours in $(k, \sigma_u)$ space for each sea state. The black horizontal line indicates $\sigma_u = 0.1$ m/s. The blue region ($\Gamma > 1$) covers the representative range of $\sigma_u$ explored here while the red region ($\Gamma < 1$) appears primarily at comparatively low turbulence levels  ($\sigma_u < 0.03$ m/s).  
All calculations are based on the JONSWAP spectrum and the wave-action based steepness formulation $\epsilon_{\mu_k} = k\sqrt{E_k \Delta k/\omega_k}$, with fixed correlation time $\tau_c = 600$ s.
}
\label{fig:multiregime}
\end{figure}
The dependence upon wavenumber is analysed in
Figure~\ref{fig:multiregime}. For every sea state investigated,
$\Gamma(k)$ exhibits a minimum close to the spectral peak.
This behaviour reflects the competition between two opposite trends. The transport rate increases approximately as
$\tau_{tr}^{-1}\sim k^2,$ whereas the resonant interaction rate is controlled primarily by the wave
steepness, $\tau_{HZ}^{-1}\sim\epsilon_\mu^4\omega_k.$

Near the spectral peak,
the increase of steepness enhances nonlinear interactions,
leading to the observed minimum.
Away from the peak,
the rapid decrease of steepness causes $\Gamma$ to increase,
particularly for long swell where resonant interactions become weak.

The time-scale comparison shown in
Figure~\ref{fig:multiregime}c
illustrates the resulting separation.
For the representative extreme-storm case,
transport acts over time scales of a few hours,
whereas resonant interactions require several days or longer over most of
the spectrum.

\subsection{Physical interpretation}
The numerical results reveal a simple physical picture. Resonant four-wave interactions are intrinsically weak,
their characteristic rate varying as the fourth power of the wave steepness. Conversely, stochastic transport depends primarily upon the intensity and persistence
of unresolved velocity fluctuations.

Because energetic oceanic flows continuously contain internal waves, submesoscale motions and turbulent currents,
the resulting transport intensity
$\sigma_u^2\tau_c$ is sufficiently large that the associated transfer rate generally exceeds the resonant interaction rate over the parameter range explored here.

This does not imply that resonant interactions become negligible. Rather, the present analysis suggests that stochastic transport may provide a comparable -- or even dominant -- mechanism for redistributing wave action under
representative ocean conditions. Both processes are expected to coexist, their relative importance being determined by the dimensionless parameter $\Gamma$.
\subsection{Implications}
These results suggest several consequences for ocean-wave modelling.The comparison indicates that unresolved-current scattering may contribute
significantly to spectral evolution alongside resonant four-wave interactions. Transport-induced redistribution offers a possible mechanism contributing to the relatively rapid swell evolution observed in the open ocean. In particular, the long-standing problem of underpredicted swell decay may admit a complementary interpretation through scattering by the ocean's turbulent velocity field, which can redistribute energy across scales and directions. Finally, the present framework predicts enhanced transport effects in regions of strong internal-wave activity,
intense submesoscale turbulence,
or energetic boundary currents,
providing several observational and numerical tests for future work.

\section{Conclusion}
We have developed a kinetic description for a coupled multiscale wave system in which large-scale linear surface waves interact with unresolved stochastic fluctuations transported by the resolved flow. Starting from a coupled multiscale stochastic water-wave system, we derived a closed kinetic equation containing two distinct mechanisms: a conservative scattering operator, analogous to wave propagation in random media, and an effective quartic interaction term generated by the transport of unresolved correlation modes. Unlike the classical Hasselmann--Zakharov collision operator, which originates from intrinsic nonlinear wave interactions, the quartic contribution derived here emerges entirely from stochastic multiscale transport.

The resulting kinetic equation exhibits two asymptotic regimes. The scattering operator gives rise to a diffusive redistribution of wave action in wavenumber space, while the quartic transport term produces nonlocal production--dissipation mechanisms and cascade dynamics through its coupling with the transported unresolved spectrum. Depending on the statistical properties of the unresolved fluctuations, the model predicts inverse transfers of wave action together with direct transfers of energy, thereby recovering cascade behaviours analogous to those of weak turbulence, although generated by a fundamentally different physical mechanism.

A numerical comparison with the classical Hasselmann--Zakharov transfer rate was then carried out using representative deep-water sea states. Introducing the dimensionless ratio
$\Gamma=\tau_{HZ}/\tau_{tr}$, allowed us to compare the characteristic transfer time scales associated with resonant four-wave interactions and stochastic transport. Over the representative parameter range explored here, $\Gamma$ generally exceeds unity, often by a substantial margin, indicating that unresolved-current-induced transport may compete with, and frequently exceed, the characteristic Hasselmann transfer rate.

This behaviour results from three complementary effects: the persistent scattering generated by unresolved currents, the relatively weak quartic nonlinear transfer associated with realistic wave steepness, and the favourable $k^2$ scaling of the stochastic transport rate. Together these mechanisms suggest that unresolved stochastic transport may constitute an important contributor to spectral evolution alongside resonant nonlinear interactions.

These results motivate a broader view of wave modelling. Rather than considering unresolved currents solely as perturbations of wave dynamics, they suggest treating stochastic transport as an active mechanism capable of generating spectral redistribution, cascade dynamics, and wave growth. This opens the possibility of extending operational spectral wave models by incorporating transport-induced scattering together with classical resonant interaction source terms. The framework also offers a possible explanation for rapid swell evolution and predicts enhanced transport effects in regions of intense internal-wave activity, submesoscale turbulence, and energetic boundary currents.

More generally, the present work suggests that stochastic transport is not merely a correction to weak turbulence theory, but provides an alternative kinetic pathway through which multiscale ocean dynamics can shape wave spectra. Although the comparison with the Hasselmann equation is based on asymptotic transfer-rate ordering rather than on a direct equivalence between collision operators, it provides a first step toward a unified stochastic kinetic theory integrating resonant wave interactions and transport by unresolved oceanic motions.

\begin{appendix}

\section{The JONSWAP Spectrum}
\label{app:jonswap}

\subsection{Foundation and Formulation}

The Joint North Sea Wave Project (JONSWAP) remains one of the most comprehensive field experiments in ocean wave physics. Conducted in 1968-1969 along a 160 km profile extending westward from the island of Sylt into the North Sea, the experiment involved up to thirteen simultaneous wave stations measuring spectra under varying fetch conditions \cite{hasselmann1973}. The resulting spectral formulation has become a cornerstone of ocean engineering and wave modeling \cite{komen1994, young1999}.

\subsection{Spectral Formulation}

The JONSWAP spectrum is an extension of the Pierson-Moskowitz spectrum for fetch-limited, developing seas \cite{pierson1964}. It is expressed as:

\begin{equation}
S(\omega) = \alpha g^2 \omega^{-5} \exp\left[-\frac{5}{4}\left(\frac{\omega_p}{\omega}\right)^4\right] \gamma^{\exp\left[-\frac{(\omega-\omega_p)^2}{2\sigma^2\omega_p^2}\right]}
\end{equation}

where $\omega$ is angular frequency, $\omega_p$ is the peak spectral frequency, $g$ is gravitational acceleration, and $\alpha$ is the Phillips constant scaling the equilibrium range \cite{phillips1958}.

The key parameters are:

\begin{itemize}
    \item \textbf{Peakedness parameter $\gamma$}: Typically averaging $\gamma = 3.3$ for the North Sea data, this factor controls the enhancement of the spectral peak relative to the Pierson-Moskowitz form \cite{hasselmann1976}. Values range from 1 to 7 depending on sea state development.
    
    \item \textbf{Spectral width parameter $\sigma$}: Accounts for the asymmetry of the peak, with:
    \begin{equation}
    \sigma = \begin{cases} 
    0.07 & \text{for } \omega \leq \omega_p \\
    0.09 & \text{for } \omega > \omega_p 
    \end{cases}
    \end{equation}
    This formulation reflects the steeper low-frequency face and gentler high-frequency decay observed in measured spectra \cite{young1999}.
    
    \item \textbf{Phillips constant $\alpha$}: Related to fetch and wind speed by:
    \begin{equation}
    \alpha = 0.076 \left(\frac{U_{10}^2}{g F}\right)^{0.22}
    \end{equation}
    where $U_{10}$ is the wind speed at 10 m height and $F$ is the fetch length \cite{hasselmann1973}.
\end{itemize}

\subsection{Physical Significance}

The JONSWAP spectrum captures several fundamental aspects of wind-wave generation \cite{komen1994}:

\begin{enumerate}
    \item \textbf{Fetch-limited growth}: Unlike the fully-developed Pierson-Moskowitz spectrum, JONSWAP explicitly accounts for the distance over which wind blows, making it applicable to coastal and regional seas.
    
    \item \textbf{Nonlinear interactions}: The characteristic overshoot of the spectral peak and its downshift with increasing fetch were explained by Hasselman et al \cite{hasselmann1973} as resulting from resonant four-wave interactions  \cite{hasselmann1962}.
    
    \item \textbf{Energy balance}: The JONSWAP measurements revealed that approximately 80-90\% of the momentum transferred from wind to waves is subsequently transferred to currents via nonlinear interactions and dissipation \cite{hasselmann1973}.
\end{enumerate}

\subsection{Parameterization for This Study}

For the present analysis, we adopt the widely-used parameterization in terms of significant wave height $H_s$ and peak period $T_p = 2\pi/\omega_p$, with the standard peakedness $\gamma = 3.3$ \cite{goda2000, young1999}. The spectrum is computed as:
\begin{equation}
S_H(f) = \alpha H_s^2 T_p \left(\frac{f}{f_p}\right)^{-5} \exp\left[-1.25\left(\frac{f}{f_p}\right)^{-4}\right] \gamma^{\exp\left[-\frac{(f-f_p)^2}{2\sigma^2 f_p^2}\right]},
\end{equation}
where the normalization factor $\alpha$ ensures consistency with the specified $H_s$ \cite{goda2000}. Following Goda's approximation \cite{goda2000}, this formulation provides a robust basis for deriving wave steepness across the frequency range.

The steepness distribution $\epsilon_\mu(k)$ follows from the wave-action based relation:
\begin{equation}
\epsilon_{\mu_k} = k\sqrt{\frac{E_k \Delta k}{\omega_k}},
\end{equation}
where $E_k = S_H(f) \cdot \frac{\dif f}{\dif k} = S_H(f) \cdot \frac{c_g}{2\pi} $, with $c_g$ the group velocity, converts the frequency spectrum to wavenumber space, and $\Delta k$ accounts for the discrete mode spacing. This formulation, grounded in wave action conservation, provides a consistent link between the observed spectral shape and the local wave steepness that governs nonlinear interaction rates.
\section{Functional Martingale central limit theorem (7.1.4) \cite{Ethier-Kurtz-86}}
\label{FCLT}
Let
\[
M^n=(M^{n,1},\ldots,M^{n,d})
\]
be a sequence of \(\mathbb R^d\)-valued local martingales, with
\(M^n_0\Rightarrow 0\). Let
\[
C(t)=\bigl(C^{ij}(t)\bigr)_{1\leq i,j\leq d},
\qquad t\geq0,
\]
be a deterministic continuous matrix-valued function such that
\(C(0)=0\) and \(C(t)-C(s)\) is non-negative definite whenever
\(t\geq s\).

Assume that one of the following two sets of conditions holds.

\medskip
\noindent
\textnormal{(a) Optional quadratic-covariation formulation.}
For every \(T>0\),
\[
\Exp\left[
\sup_{0\leq t\leq T}
\left|\Delta M^n_t\right|
\right]
\longrightarrow0,
\]
and, for every (i,j) and \(t\geq0\),
\[
[M^{n,i},M^{n,j}]_t
\longrightarrow
C^{ij}(t)
\]
in probability.

\medskip
\noindent
\textnormal{(b) Predictable quadratic-covariation formulation.}
For every (T>0),
\[
\Exp\left[
\sup_{0\leq t\leq T}
\left|\Delta M^n_t\right|^2
\right]
\longrightarrow0,
\]
and, for every (i,j) and \(t\geq0\),
\[
\left\langle M^{n,i},M^{n,j}\right\rangle_t
\longrightarrow
C^{ij}(t)
\]
in probability.

Then
\[
M^n\Longrightarrow M
\qquad\text{in}\qquad
D([0,\infty),\mathbb R^d),
\]
where \(D\) is the Skorokhod space of c\`{a}dl\`{a}g functions (right continuous with left limits) and  \(M\) is a continuous centered Gaussian process with independent
increments and covariance
\[
\Exp\left[M(t)M(t)^{\mathsf T}\right]
=
C(t).
\]

\end{appendix}
\bibliographystyle{plain}
\bibliography{biblio}
\end{document}